\documentclass[twocolumn]{aastex62}

\begin{document}

\title{Light Element Abundances and Multiple Populations in M10}

\author{Jeffrey M. Gerber}
\affiliation{Astronomy Department, Indiana University Bloomington, Swain West 319, 727 East 3rd Street, Bloomington, IN 47405-7105, USA}
\email{jemigerb@indiana.edu}

\author{Eileen D. Friel}
\affiliation{Astronomy Department, Indiana University Bloomington, Swain West 319, 727 East 3rd Street, Bloomington, IN 47405-7105, USA}
\email{efriel@indiana.edu}
\author{Enrico Vesperini}
\affiliation{Astronomy Department, Indiana University Bloomington, Swain West 319, 727 East 3rd Street, Bloomington, IN 47405-7105, USA}
\email{evesperi@indiana.edu}

\begin{abstract}
We present CN and CH band measurements for 137 RGB and AGB stars in the Galactic globular cluster M10. Our measurements come from low resolution spectroscopy taken with the Hydra spectrograph on the WIYN-3.5m telescope. We use these measurements to identify two populations of stars within the cluster, a CN-normal and CN-enhanced, and find that in our sample 60\% of stars are CN-enhanced. Our large sample allows us to conduct a detailed analysis on the carbon and nitrogen abundances and the radial distribution of each population separately. Our analysis of the radial dependence shows that each population has the same radial distribution in the cluster, which is likely due to the cluster being dynamically evolved. We also compare our results to other methods of classifying multiple populations in globular clusters such as the Na-O anti-correlation and the HST pseudo-color magnitude diagrams. We find that these three methods of identifying multiple populations are in good agreement with each other for M10 and all lead to an estimate of the fraction of second generation stars approximately equal to 60\%. Among AGB stars, when classified by the CN band, there appears to be a lack of second generation stars when compared to the RGB stars. However, when classified by [N/Fe], we find a similar 60\% of AGB stars in the second generation. Finally, we use the measured carbon and nitrogen abundances in RGB stars to study the change of each element with magnitude as stars evolve up the RGB, comparing the results to globular clusters of similar metallicity, M3 and M13. 
\end{abstract}


\section{Introduction}
Until recently, Galactic globular clusters (GC) were considered simple stellar populations formed out of the same material and therefore characterized by the same chemical composition. However, numerous photometric and spectroscopic studies have shown that the stellar populations in GCs are far from simple and provided strong evidence that these systems host multiple stellar populations characterized by a spread in light elements and anti-correlations between C-N, Na-O, Mg-Al (e.g., \citealt[and references therein]{gratton}). 

The first studies that discovered light element abundance variations were low resolution spectroscopic studies focused on the blue CN and CH bands around 3800 \AA~and 4300 \AA, respectively \citep[e.g.,][and references therein]{suntzeff,kraft}. Differences in these bands among the brightest red giant branch (RGB) stars indicated variations in carbon and nitrogen along the RGB. What was still unknown, however, was whether or not these variations were primordial or caused by evolutionary effects that had taken place as the stars ascended the RGB. At first, the primordial solution seemed unlikely due to the homogeneous nature of the heavier elements in the clusters.

We now know that both primordial abundance variations and evolutionary effects are behind these light element variations in the RGB stars of GCs. Studies over the last twenty years have shown that the variations in CN and CH band strength (and therefore carbon and nitrogen abundance) exist even on the main sequence and sub-giant branch of GCs, which means the inhomogeneities are primordial \citep[e.g.,][etc.]{2dfrutti,briley1992,briley2004b,cohena,cohen2005}. The observed chemical properties of the multiple populations in GCs suggest that a significant fraction of stars in these systems formed out of gas processed in a previous generation of polluting stars. A number of different polluters have been proposed in the literature including massive AGB stars \citep[see e.g.,][]{ventura2001,dantona2016}, fast rotating massive stars \citep{maeder2006,prantzos2006,decressin2007a,decressin2007b}, massive binary stars \citep{demink2009}, and supermassive stars \citep{denissenkov2014}; but the origin of the gas out of which the different stellar populations in GCs formed is still debated and a matter of intensive investigation.

However, evolutionary effects are also evident in data that show the carbon abundance of RGB stars decreasing with increasing luminosity. This decrease begins after the stars have passed the luminosity function bump (LFB), which is an evolutionary ``stall" seen on the RGB \citep[e.g.,][and references therein]{suntzeff,suntzeff91,m13cfe}.

The stall is caused when the H-burning shell around the core of a low-mass RGB star advances outwards and encounters a large difference in the mean molecular weight, called the $\mu$-barrier. The $\mu$-barrier is created (or left behind) by the convective envelope at its deepest penetration into the star during the first dredge up \citep{iben1965}. As the H-burning shell expands, it encounters the $\mu$-barrier and the sudden influx of hydrogen-rich material causes the star to become bluer and fainter. The star then reaches equilibrium and continues to evolve up the RGB \citep{iben1968,cassisi2002}. In the color-magnitude diagrams (CMDs) of populous GCs, this evolutionary stutter is observed as a large number of stars all at the same magnitude, which creates the LFB.

The astrophysical cause of this evolutionary depletion in carbon and simultaneous increase in nitrogen has been the subject of debate. While lower carbon isotope ratios in late RGB stars indicate that this material comes from CN(O)-cycle processing deep within the star, the question is what mechanism is allowing it to move past the radiative zone which separates the hydrogen burning envelope from the convective layer and should prevent any material from passing between the two. Some of the theories describing the cause of this deep mixing, which is distinct from the first and second dredge events and only appears to occur in stars near solar mass (0.5-2.0 $M_\odot$), are rotational mixing \citep{sweigart,chaname,palacios}, magnetic fields \citep{palmerini,nordhaus,busso,hubbard}, internal gravity waves \citep{denissenkov2000}, and thermohaline mixing \citep{eggleton2006,eggleton2008,charbonnel}. While the latter of these mechanisms seems to be the most promising, it still requires more data to better understand and constrain the theory.

While previous studies using spectroscopy to measure C and N have identified multiple populations and the characteristics of deep mixing, they have often lacked the sample sizes necessary to model these properties with statistical significance. The logical next step is to increase the sample size over previous studies to better constrain models of the underlying phenomena creating the observed light element inhomogeneities.
We present our study of low resolution spectroscopy covering the blue CN and CH bands for a uniform sample of over 120 RGB stars in M10. Our study has an advantage over previous studies of M10 due to its much larger sample size, which allows us to track the C, N abundances of the stars in each population as they climb the RGB. In addition, the broad radial coverage of our data allows us to study the spatial distribution of the multiple populations separated by CN band strength.

We chose to study M10 for a number of reasons. First, its C and N abundances have not been well studied previously; \citet{smith} studied C and N for only 15 stars. However, Na and O have been determined by \citet{gir,uves} for numerous stars in the cluster, so classification of populations by different abundance indicators can be made. Second, at an intermediate metallicity of [Fe/H] $\sim$ -1.5, M10 also provides an interesting comparison with other well studied clusters such as M3 and M13. Lastly, M10 provides a midpoint in metallicity as part of a larger study we are undertaking of GCs covering a range of metallicities from [Fe/H] $\sim$~-2 to -0.7. 

In the following paper we will present our full sample of 124 RGB and 13 AGB stars. Section \ref{Observations} describes our selection criteria and the data reduction. Section \ref{Analysis} discusses the band measurements and calculations of C and N abundances for all stars in the sample. A discussion of the spatial distribution for the separate populations of stars in M10, the evolution of C and N with magnitude, and comparison to other clusters of similar metallicity (namely, M13 and M3) are in Section \ref{Results}. We summarize our results and draw conclusions in Section \ref{Conclusions}.

\section{Observations and Data Reduction} \label{Observations}

\subsection{Observations and Target Selection}
We obtained 190 different spectra of stars within a 18' by 18' grid around the center of M10 in two observation runs from 1-4 Aug. 2014 and 10-12 Jun. 2016 using the Wisconsin-Indiana-Yale-NOAO (WIYN)\footnote{The WIYN Observatory is a joint facility of the University of Wisconsin-Madison, Indiana University, the National Optical Astronomy Observatory and the University of Missouri} 3.5m Telescope and Hydra, a multi-object, fiber-fed bench spectrograph. The Bench Spectrograph was used with the ``600@10.1" grating, which resulted in spectra with a $\sim4.5$~\AA~pixel$^{-1}$ dispersion covering a range of $\sim2800$~\AA. The spectra taken during the 2014 run are centered at a wavelength of $4900$~\AA, while the spectra taken during the 2016 run are centered at $5100$~\AA.
Six different configurations of fibers were necessary to obtain the full sample size.

\begin{figure}
\centering
\includegraphics[trim = 0.4cm 0.4cm 0.4cm 0.4cm, scale=0.43, clip=True]{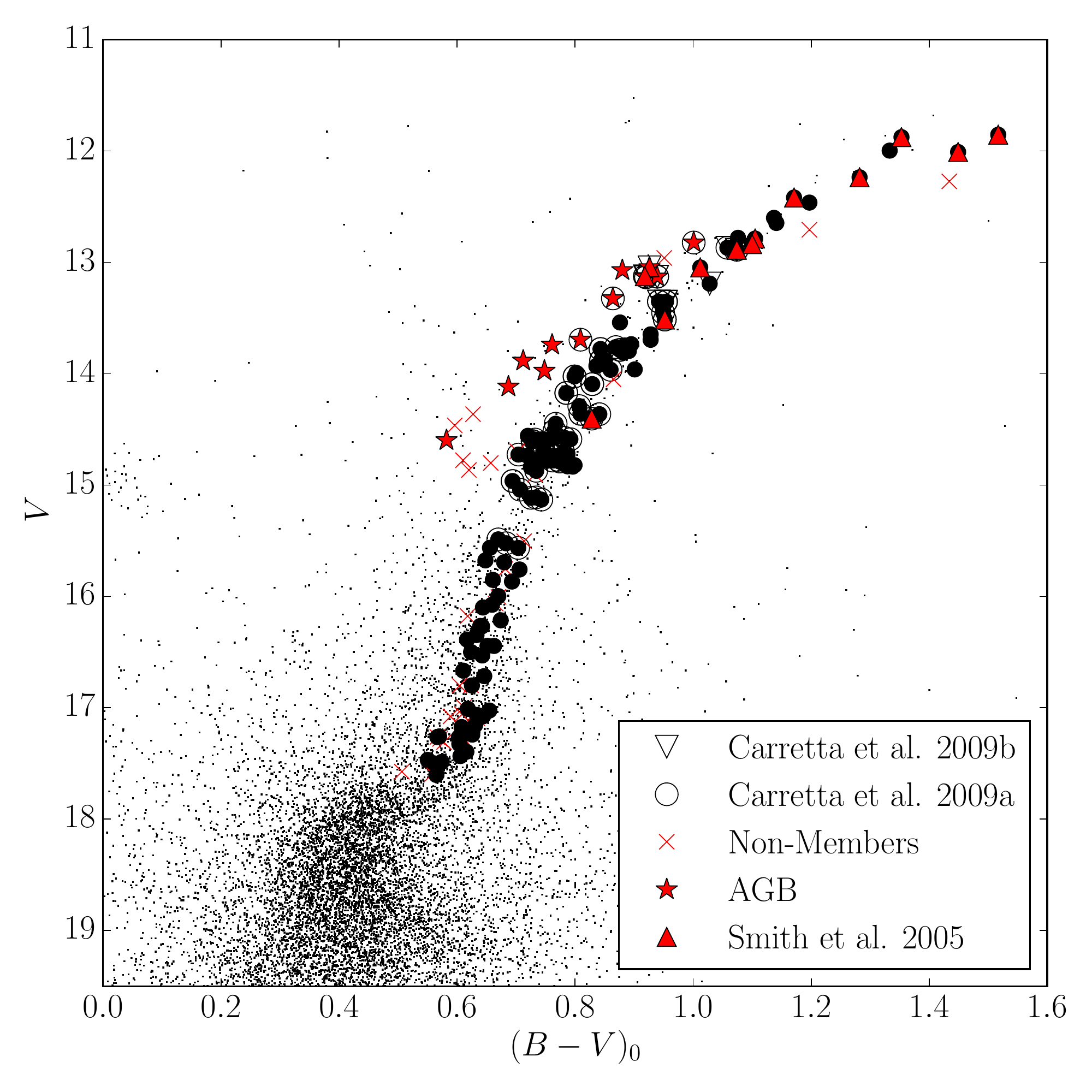}
\caption{Observed stars are shown in the color-magnitude diagram taken from \citet{photometry}. RGB stars are indicated by filled circles and AGB stars are shown as red stars. Stars observed by \citet{gir, uves} and \citet{smith} are also indicated as shown in the key above. Non-members as determined by their radial velocities are shown as red x's.}
\label{cmd-sample}
\end{figure}

We selected our target stars based on their location in the $V$ versus $(B-V)$ CMD using photometry from \citet{photometry}. Stars were chosen to have $V \leq 17.5$, which allows us to cover the entire RGB of M10 down to the sub-giant branch as shown in Figure \ref{cmd-sample}. We also selected stars covering a large range of radii from the cluster center, allowing us to study trends with both magnitude and spatial distribution in the multiple populations in M10.

Another criterion for stars in our sample was whether or not the stars had measured Na and O abundances in the literature, so that we could compare our results separating the populations using the C-N anti-correlation to those using the Na-O anti-correlation. Stars with Na and O literature values from either \citet{gir}, \citet{uves}, or \cite{smith} were given high priority. We also included 13 stars from \citet{smith} in our sample to allow comparison between our values for C and N. Figure \ref{cmd-sample} shows which stars in our sample overlap with either \citet{gir}, \citet{uves}, or \citet{smith}. Some stars were measured on multiple nights to allow for a statistical determination of our uncertainties.

\subsection{Data Reduction and Radial Velocities} \label{datareduction}

We reduced the data with the standard IRAF\footnote{IRAF is distributed by the National Optical Astronomy Observatories, which are operated by the Association of Universities for Research in Astronomy, Inc., under cooperative agreement with the National Science Foundation.} software to perform bias subtraction. The package \textit{dohydra} was then used for flat fielding, dispersion correction with a CuAr comparison lamp spectrum, and extraction to one-dimensional spectra. We exposed configurations of bright stars (with $M_v \leq 0.5$) for a total integration time of $\sim$1 hr over 7-10 exposures and configurations of faint stars (with $M_v > 0.5$) for a total integration time of $\sim$2.5-3 hrs over 5-6 exposures to reduce the effects of cosmic rays and prevent saturation of the CCD. The processed spectra from each exposure were combined using the IRAF task \textit{scombine}. Final combined spectra had a S/N of $\sim$20-40 at 3883 \AA~(the CN band) and $\sim$100 at 4300 \AA~(the CH band).

Membership was determined for the stars in our sample using radial velocities measured with the IRAF task \textit{fxcor}. Stars were cross correlated against a radial velocity standard, which was observed on multiple nights during the 2016 run to ensure a consistent velocity measurement. To be certain that the standard star could be used to determine accurate velocities for the 2014 run, the radial velocity of the standard star was measured against a sky flat taken during the 2014 run. Since the correct velocity for the standard was determined with this method, we decided the standard star would be effective at measuring accurate velocities for both runs.

We determined a median velocity of 71 km s$^{-1}$ with a standard deviation of 11 km s$^{-1}$ for the cluster based on our observations. Individual measurement uncertainties from \textit{fxcor} are $\sim$20 km s$^{-1}$, while differences between multiple measurements for 23 stars have standard deviations of ~10-15 km s$^{-1}$. Even with our large uncertainties, we find good agreement with values in the literature. \citet{dinescu} found a radial velocity of 75.5 $\pm$ 1.1 km s$^{-1}$ and \citet{gir,uves} found 73.92 $\pm$ 4.95 km s$^{-1}$ from higher resolution data. 

We consider stars falling outside $3\sigma$ of our median velocity for the cluster to be non-members. Without these non-members, we are left with 137 stars including 123 RGB stars and 14 AGB stars. These stars are distinguished in Figure \ref{cmd-sample}.

\section{Analysis} \label{Analysis}

\subsection{CN and CH Bands}
\subsubsection{Index Definitions}

Measuring the molecular band strength of a given spectrum typically involves taking the ratio of the integrated flux of a wavelength range containing the molecular feature to the integrated flux of a nearby area of the continuum. A number of CN and CH bands have been defined over the years to optimize band measurements in various luminosity and metallicity ranges \citep[see][]{martell}. For the CN band, we chose to use the S(3839) band from \citet{bands} to avoid contamination of hydrogen lines that become present in stars lower on the RGB.
We also use the CH(4300) index defined by \citet{bands} to allow easy comparison between our work and the literature. We weight our CH bandpass integrations by the width in angstroms of the spectral bandpass window. Our final band definitions are:

\begin{equation}
S(3839) = - 2.5log\frac{F_{3861-3884}}{F_{3894-3910}}
\end{equation}
\begin{equation}
CH(4300) = -2.5log\frac{F_{4285-4315}/30}{0.5(C_1 + C_2)}
\end{equation}
\begin{equation}
C_1 = F_{4240-4280}/40
\end{equation}
\begin{equation}
C_2 = F_{4390-4460}/70
\end{equation}

We used a trapezoid rule with non-uniform step sizes to numerically integrate these windows and calculate the bands. We note that \citet{boberg6791} points out how the numerical method of calculating these integrals can affect the final band measurements, especially in spectra with a S/N below 40. However, these effects are lower than the uncertainty in our bands caused by other factors such as night-to-night variations, which we can evaluate from stars measured on multiple nights.

\subsubsection{Flux Calibration} \label{flux_calibration}
Because the S(3839) band only has one comparison window, the measurement can be affected by small changes in the continuum. Therefore, it was essential to correct for the instrumental response, so that the shape of the continuum better matched what would be expected for the surface conditions of the stars observed. 

To fit and remove the instrumental response, we followed a similar method to that of \citet{2dfrutti}. We first generated appropriate synthetic spectra for each star based on their atmospheric parameters such as effective temperature and surface gravity. Atmospheric parameters were determined by the infrared flux method as outlined in \citet{alonso,alonsoerrat}, which makes use of V-K colors to determine an effective temperature. K magnitudes were from the 2MASS survey \citep{2mass}, and V magnitudes were from \citet{photometry}. For stars without a K magnitude measured by 2MASS, we used the relationship between B-V colors and temperature from \citet{alonso}. Magnitudes were transformed to the TCS photometric system following the relations summarized in \citet{colortransforms}, and we adopted an E(B-V) of 0.28 \citep{photometry}. Once effective temperatures were calculated, we then calculated surface gravities using the bolometric corrections given by \citet{alonso}. We also transformed the apparent magnitudes to absolute magnitudes using the apparent distance modulus (m -- M)$_{v}$ = 14.18 \citep{photometry}. Our final temperatures, surface gravities, and absolute magnitudes are listed in Table \ref{tab:alldata}, which can be found in its entirety online.

\begin{deluxetable*}{cccccccccccccccccc}
\tabletypesize{\scriptsize}
\tablecolumns{18}
\tablewidth{0pt}
\tablecaption{Stars Measured in M10
\label{tab:alldata}}
\tablehead{
\colhead{ID$^{1}$} & \colhead{RA$^{1}$} & \colhead{Dec$^{1}$} & \colhead{RV$_{hel}$} & \colhead{$V^{1}$} & \colhead{M$_v^{2}$} & \colhead{$K_{2MASS}$} & \colhead{$T_{eff}$} & \colhead{logg} & \colhead{CN} & \colhead{$\delta$CN} & \colhead{CH} & \colhead{[C/Fe]} & \colhead{[N/Fe]} & \colhead{[O/Fe]$^{3,4}$} & \colhead{[Na/Fe]$^{4}$} & \colhead{Branch} & \colhead{Memb} \\ 
& \colhead{(J2000)} & \colhead{(J2000)} & \colhead{(km s$^{-1}$)} & & & & & & & & & & & & &}
\rotate
\startdata
    9 &  254.29484 & -4.07889 &   81.0 &  11.85 & -2.33 &      7.62 &  3964.0 &  0.68 & -0.062 &    0.087 &  0.325 &   -1.00 &    1.59 &  -0.110 &    \nodata &    RGB &    y \\
   11 &  254.25159 & -4.10039 &   68.0 &  11.88 & -2.30 &      7.90 &  4098.0 &  0.81 & -0.203 &   -0.059 &  0.354 &   -0.56 &    0.77 &   0.230 &    \nodata &    RGB &    y \\
   14 &  254.38488 & -4.05021 &   62.0 &  12.00 & -2.18 &      8.06 &  4120.0 &  0.88 & -0.043 &    0.077 &  0.333 &   -0.72 &    1.22 &   \nodata &    \nodata &    RGB &    y \\
   15 &  254.28968 & -4.12285 &   91.0 &  12.01 & -2.17 &      7.84 &  3994.0 &  0.77 &  0.008 &    0.126 &  0.311 &   -1.13 &    1.77 &  -0.260 &    \nodata &    RGB &    y \\
   22 &  254.28057 & -4.12394 &   80.0 &  12.24 & -1.94 &      8.38 &  4165.0 &  1.01 &  0.100 &    0.189 &  0.306 &   -0.84 &    1.68 &   \nodata &    \nodata &    RGB &    y \\
   23 &  254.26598 & -3.99828 & -123.0 &  12.27 & -1.91 &      7.65 &  3800.0 &  0.67 & -0.136 &   -0.050 &  0.367 &   -0.52 &    1.87 &   \nodata &    \nodata &    RGB &    n \\
   26 &  254.27816 & -4.05295 &   77.0 &  12.42 & -1.76 &      8.78 &  4310.0 &  1.19 &  0.159 &    0.236 &  0.298 &   -0.78 &    1.62 &   \nodata &    \nodata &    RGB &    y \\
   28 &  254.32108 & -4.22408 &   66.0 &  12.46 & -1.72 &      8.76 &  4267.0 &  1.17 &  0.224 &    0.299 &  0.316 &   -0.72 &    1.70 &   \nodata &    \nodata &    RGB &    y \\
   33 &  254.39517 & -4.11894 &   75.0 &  12.60 & -1.58 &      8.95 &  4299.0 &  1.25 & -0.120 &   -0.047 &  0.361 &   -0.36 &    0.82 &   \nodata &    \nodata &    RGB &    y \\
   36 &  254.31011 & -4.18646 &   67.0 &  12.65 & -1.53 &      9.00 &  4301.0 &  1.27 & -0.209 &   -0.135 &  0.363 &   -0.36 &    0.65 &   \nodata &    \nodata &    RGB &    y \\
   39 &  254.43930 & -4.18652 &  -82.0 &  12.71 & -1.47 &      8.98 &  4250.0 &  1.26 & -0.181 &   -0.107 &  0.362 &   -0.37 &    0.73 &   \nodata &    \nodata &    RGB &    n \\
   43 &  254.26157 & -4.17248 &   80.0 &  12.78 & -1.40 &      9.28 &  4413.0 &  1.40 & -0.175 &   -0.099 &  0.343 &   -0.40 &    0.77 &   \nodata &    \nodata &    RGB &    y \\
   45 &  254.26936 & -4.10946 &   78.0 &  12.79 & -1.39 &      9.23 &  4364.0 &  1.37 &  0.003 &    0.080 &  0.316 &   -0.73 &    1.15 &  -0.070 &    \nodata &    RGB &    y \\
   47 &  254.29493 & -4.11179 &   59.0 &  12.82 & -1.36 &      9.52 &  4560.0 &  1.51 & -0.192 &   -0.114 &  0.290 &   -0.53 &    0.93 &   \nodata &   -0.001 &    AGB &    y \\
   49 &  254.30720 & -4.14313 &   77.0 &  12.84 & -1.34 &      9.24 &  4342.0 &  1.37 & -0.137 &   -0.059 &  0.340 &   -0.43 &    0.87 &   \nodata &    \nodata &    RGB &    y \\
   51 &  254.30114 & -4.09260 &   63.0 &  12.87 & -1.31 &      9.54 &  4460.0 &  1.46 & -0.178 &   -0.098 &  0.331 &   -0.43 &    0.80 &   0.373 &   -0.044 &    RGB &    y \\
   53 &  254.27298 & -4.14345 &   80.0 &  12.89 & -1.29 &      9.39 &  4413.0 &  1.44 &  0.076 &    0.156 &  0.324 &   -0.39 &    1.09 &   0.319 &    0.332 &    RGB &    y \\
   54 &  254.30902 & -4.09251 &   68.0 &  12.89 & -1.29 &      9.39 &  4410.0 &  1.44 & -0.137 &   -0.056 &  0.338 &   -0.31 &    0.90 &   0.596 &    0.003 &    RGB &    y \\
   59 &  254.15202 & -4.23894 &  108.0 &  12.96 & -1.22 &      9.68 &  4579.0 &  1.57 & -0.158 &   -0.073 &  0.353 &   -0.28 &    0.76 &   \nodata &    \nodata &    RGB &    n \\
   62 &  254.30327 & -4.05438 &   71.0 &  13.04 & -1.14 &      9.85 &  4662.0 &  1.65 & -0.146 &   -0.056 &  0.231 &   -0.74 &    1.29 &   0.303 &    0.249 &    AGB &    y \\
   63 &  254.29659 & -4.05532 &   72.0 &  13.05 & -1.13 &      9.67 &  4505.0 &  1.56 &  0.153 &    0.243 &  0.282 &   -0.82 &    1.48 &  -0.340 &    \nodata &    RGB &    y \\
   65 &  254.30013 & -4.10028 &   64.0 &  13.07 & -1.11 &      9.94 &  4733.0 &  1.70 & -0.170 &   -0.078 &  0.198 &   -0.80 &    1.42 &   \nodata &    \nodata &    AGB &    y \\
   67 &  254.28263 & -4.16026 &   68.0 &  13.12 & -1.06 &      9.94 &  4677.0 &  1.69 & -0.244 &   -0.148 &  0.270 &   -0.46 &    0.91 &   0.527 &   -0.207 &    AGB &    y \\
 \enddata
 \tablecomments{1. \citet{photometry}, 2. (m -- M)$_{v}$ = 14.18, 3. \citet{smith}, 4. \citet{gir,uves} \\ The complete table is available online.}
 \end{deluxetable*}

Because we were only concerned with determining the shape of the continuum, we assumed an [Fe/H] of -1.56 based on the value from \citealt{harris} (2010 edition), and used solar abundances for carbon, nitrogen, and oxygen \citep{asplund}. Synthetic spectra were created using the Synthetic Spectrum Generator (SSG) \citep[][and references therein]{ssg}, which makes use of MARCS model atmospheres \citep{marcs}. The continua of our observed spectra were then normalized to the continua of the theoretical spectra generated for each star. The normalization was conducted by calculating the ratio between an observed stellar spectrum's continuum and its theoretical continuum at various flux points carefully chosen to avoid regions affected by CN and CH. The ratio was then fit with a spline function, which was normalized to one. Finally, each observed spectrum was divided by its normalized fit to produce a spectrum with a continuum free of instrumental response.

\subsubsection{Band Measurements} \label{bandmeasurements}
\begin{figure}
\centering
\epsscale{2.4}
\plottwo{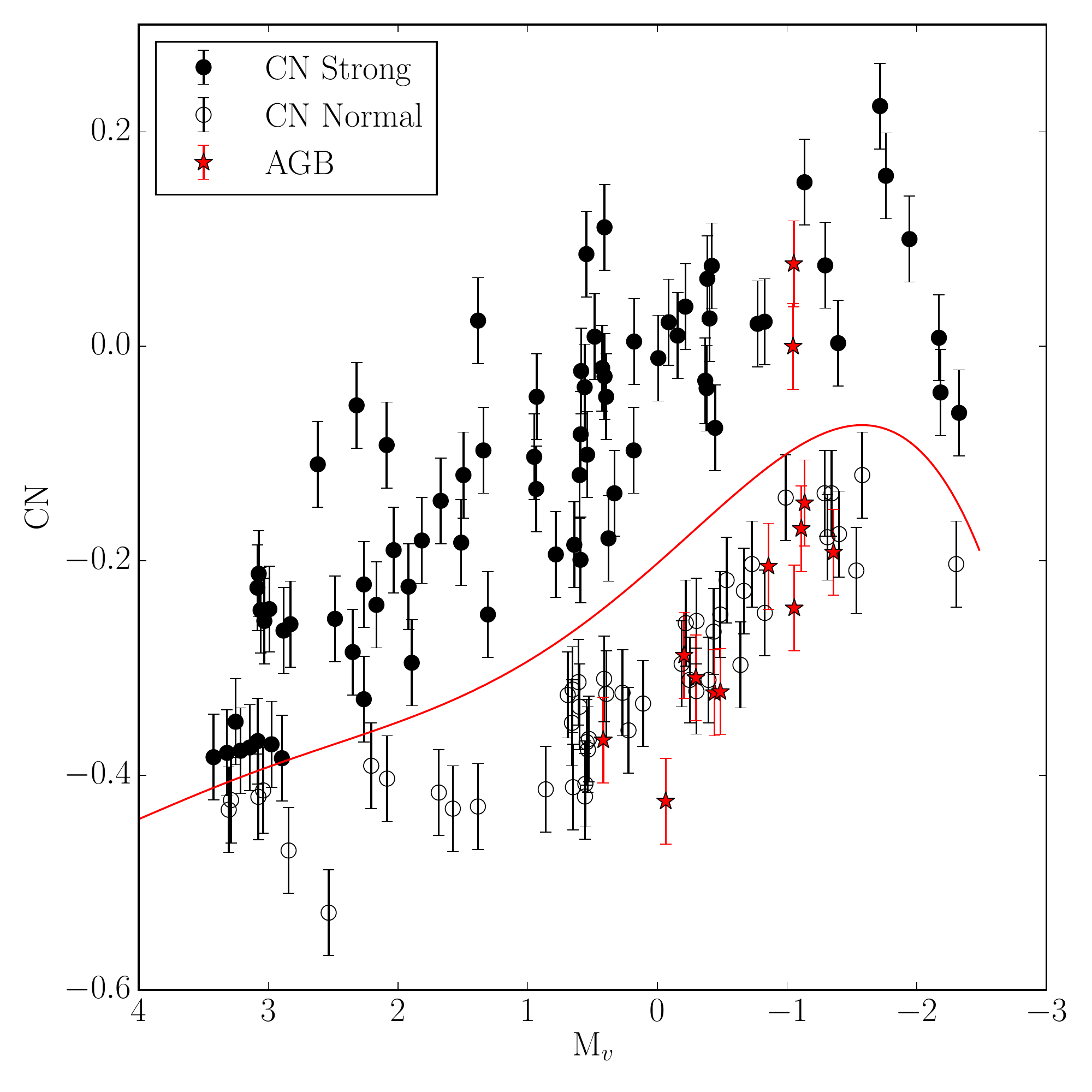}{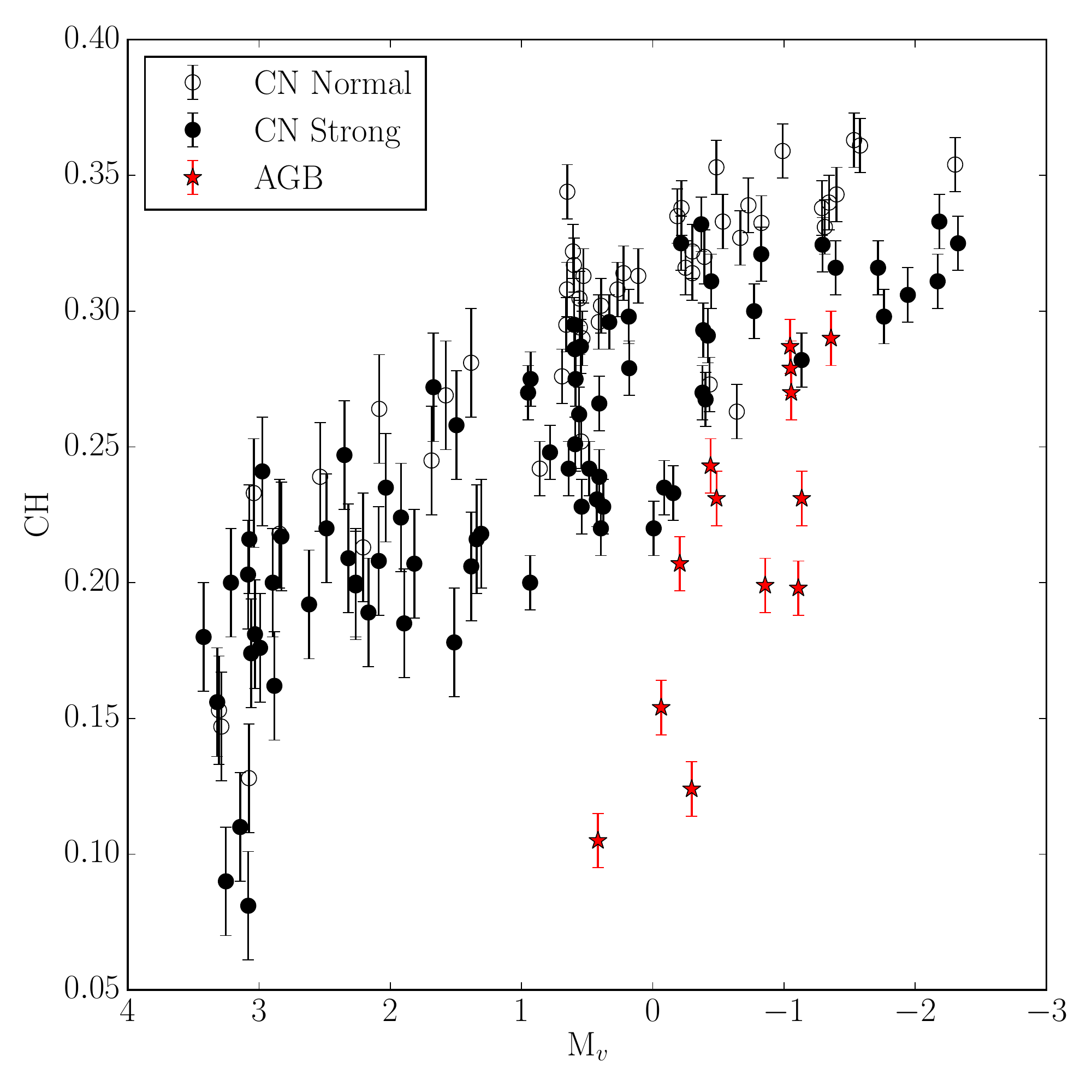}
\caption{\textbf{Top Panel:} CN band strengths as a function of magnitude for stars in our sample. The red line shows the fiducial used to determine the $\delta$CN index strength, and identify a star as CN-enhanced (filled) or CN-normal (open). AGB stars are indicated as red stars and RGB stars are shown as circles (filled or open). \textbf{Bottom Panel:} CH band strengths as a function of magnitude using the same symbols as the left panel. Both bands show a clear trend in magnitude, which is caused by surface temperature and pressure effects.}
\label{bands}
\end{figure}

CN and CH measurements were made on the flux calibrated spectra and are plotted in Figure \ref{bands} as a function of absolute magnitude, M$_v$. Our uncertainties in this figure are derived from the standard deviation of measurements for stars that were observed across either multiple nights or multiple runs. Stars were divided into two categories: bright with M$_v < 1.0$ and faint with M$_v \geq 1.0$. A median standard deviation was determined for stars in each category and applied to all stars in that category. For the CN band, both categories had similar standard deviations, so an uncertainty of 0.04 was adopted for all stars. For the CH band, this method gave an uncertainty of 0.01 for the bright stars and 0.02 for the faint stars. 

The effect of surface gravity and effective temperature on the band strength can be clearly seen in Figure \ref{bands} as both bands tend to increase with increasing luminosity. The CN band also shows a clear separation into two populations (CN-enhanced and normal), especially at the brighter magnitudes. It should also be noted that the AGB stars in the sample appear to have low CN and CH band strengths compared to the RGB stars at similar magnitude; we will discuss the behavior of AGB stars later (see Section \ref{AGB}). While dividing the two populations by eye seemed reasonable, we wanted to establish an objective way of separating the CN-enhanced and normal populations. Having an objective method for separating populations was especially important for the faint stars in the sample where the band begins to lose sensitivity, and the populations are not as well separated. 

\begin{figure*}
\centering
\includegraphics[trim = 0.4cm 0cm 0cm 0cm, scale=0.3, clip=True]{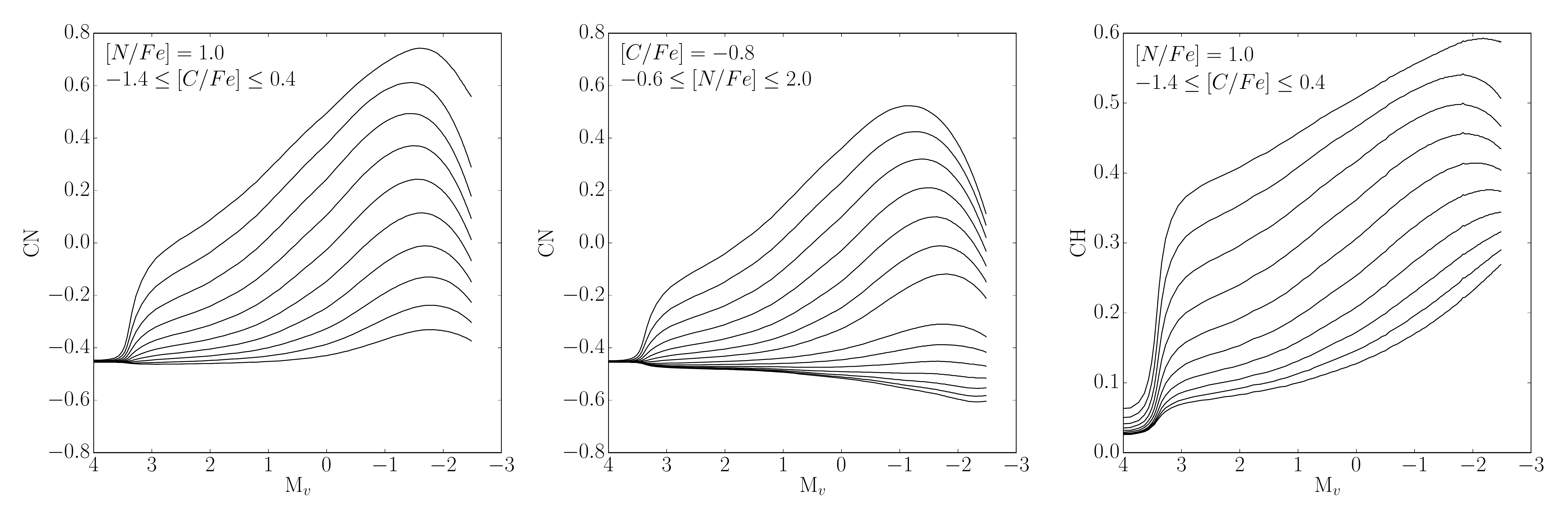}
\caption{Lines of equal abundance in the CN/CH vs. M$_v$ plane created using model spectra generated by SSG \citep{ssg}. All models used [Fe/H] = -1.56 (Harris 1996 (2010 edition)). \textbf{Left Panel:} CN as a function of magnitude with lines of constant [C/Fe] ranging from -1.4 to 0.4 dex by steps of 0.2 dex with [N/Fe] = 1.0 dex. \textbf{Middle Panel:} CN as a function of magnitude with lines of constant [N/Fe] ranging from -0.6 to 2.0 by steps of 0.2 dex with [C/Fe] = -0.8. \textbf{Right Panel:} CH as a function of magnitude with lines of constant [C/Fe] ranging from -1.4 to 0.4 dex by steps of 0.2 dex with [N/Fe] = 1.0 dex. [O/Fe] was assigned as 0.1 for [N/Fe] $>$ 0.8 and 0.4 for [N/Fe] $<$ 0.8, which were determined to be the average [O/Fe] values for N-enhanced and N-normal stars based on the data from \citet{gir,uves}.}
\label{isolines}
\end{figure*}

To create an objective way to organize the stars into populations, we first defined a $\delta$CN index by modeling and removing the effect of luminosity on the band strength. The CN band dependence on luminosity was modeled using synthetic spectra created with the SSG. Surface gravities and effective temperatures were determined using a Yonsei-Yale, \textit{YY}$^2$ isochrone \citep{yy,yi2001} generated for a 12 Gyr cluster with [$\alpha$/Fe] of 0.3 dex and [Fe/H] of -1.56. [C/Fe] was varied from -1.4 to 0.4 dex, and [N/Fe] was varied from -0.6 to 2.0 dex. [O/Fe] values were chosen based on the abundances found by \citet{gir,uves}, as explained in the next section. The final model generated band strengths are shown in Figure \ref{isolines} as lines of constant abundance (``isoabundance" lines) in the CN and CH vs. M$_v$ plane. These isoabundance lines represent how the bands change with magnitude due to changes in surface gravity and effective temperature. The trends seen with changing magnitude in the isoabundance lines also appear in the measured band strengths shown in Figure \ref{bands}. Figure \ref{isolines} also shows how the CN and CH bands lose sensitivity near the faint magnitude limit for our sample, especially at magnitudes fainter than M$_v$ $\sim$ 3.0 where the temperature of the stars is such that the molecules begin to dissociate causing the band strengths to weaken and converge.

A model was then chosen representing the average carbon and nitrogen abundances for the cluster based on the 15 stars previously measured by \citet{smith}. We also accounted for evolutionary effects that change carbon and nitrogen abundance with magnitude following the measured carbon and nitrogen abundances discussed in Section \ref{evolution}. A spline was then fit to the average isoabundance line, and used as a dividing line between the two populations as shown in Figure \ref{bands}. The $\delta$CN index was calculated by subtracting the measured CN value from the average isoabundance fit at that magnitude. All stars with a positive $\delta$CN value were identified as CN-enhanced, and those with a negative $\delta$CN were identified as CN-normal. In Figure \ref{bands}, points are coded based on their $\delta$CN strength with CN-enhanced stars indicated as filled points and CN-normal stars indicated as open.

\subsection{Determining C and N Abundances} \label{candn}
We can use the measured CN and CH bands to determine C and N abundances, following the methods of \citet{briley2004a,briley2004b}. Using the SSG, which makes use of MARCS model atmospheres \citep{marcs}, we generate synthetic spectra for each star and then vary the C and N abundances of the synthetic spectrum until it matches that star's CH and CN band measurements. We used the same effective temperatures and surface gravities determined for flux calibration (see section \ref{flux_calibration}). We assumed a microturbulence of 2.0 km s$^{-1}$ and a $C^{12}/C^{13}$ ratio of 4.0, which are reasonable for RGB stars in GCs \citep[e.g.,][]{suntzeff91,pavlenko}. For the [O/Fe] value for each star, we used the value provided by \citet{gir,uves} if available. If no value was measured by \citet{gir,uves}, we used measurements from \citet{kraft1995}. In cases where [O/Fe] was not measured for the star, we assumed an [O/Fe] of 0.11 dex for CN-enhanced stars and 0.39 dex for CN-normal stars. These abundances are based on the average values for the second and first generations, respectively, from the measurements made by \citet{gir,uves}.

To determine uncertainties in our abundance measurements, we consider separately the impact of uncertainties in the effective temperature and surface gravity, the [O/Fe] value, and the band measurements. We account for all of these factors by adjusting values in the following ways and then recalculating abundances for each star. First, we adjust the temperature by 150 K. Second, we swap the [O/Fe] of CN-enhanced and CN-normal stars. Finally, we increase every star's CN band and decrease the CH band by their respective uncertainties, which maximizes the impact of the band uncertainties since the bandstrengths are negatively correlated.\footnote{We note that the final abundances determined do not depend heavily on the microturbulence or $C^{12}/C^{13}$ with adjustments to both values giving changes on the order of hundredths of a dex.} We then evaluate how the C and N abundances would be different for each factor and combine them in quadrature. This method gives an uncertainty of $\sim$0.2-0.25 dex for the [C/Fe] and [N/Fe] of stars with M$_v <$1, and an uncertainty of 0.35 dex for the fainter stars. For the [C/Fe] values, the uncertainties of the brightest stars (M$_v < -1$) are dominated by the [O/Fe] uncertainty, which causes a change up to 0.15 dex (if no [O/Fe] is measured and the wrong average O abundance is used). However, around M$_v = -1$ and fainter the effective temperature becomes the dominating factor in the [C/Fe] uncertainty as its influence can cause a change of 0.3 dex, while the uncertainty in the O abundance only causes a change of 0.05 dex. For [N/Fe] measurements, the uncertainties are dominated by the uncertainties in the CN band strengths. After that, the effective temperature and surface gravity are the next dominant effect.

The uncertainties of the faint stars (M$_v >$1) are larger due to the following two effects. First, they have a lower S/N and therefore have higher measurement uncertainties than the bright stars. Second, the CN band loses sensitivity at faint magnitudes, as shown by the convergence of isoabundance lines in Figure \ref{isolines}, which is caused by the temperatures of these stars being hot enough to dissociate the molecules in their atmospheres. Therefore for faint stars, bands will be weaker and more difficult to measure, and less sensitive to abundance changes, leading to higher uncertainties in derived abundances. For all of these reasons, our classification into CN-enhanced and CN-normal is less secure at the faint end. The left panel of Figure \ref{bands} shows that the uncertainties for many of the faint stars in our sample could lead them to be classified in either population.

\begin{figure}
\centering
\includegraphics[trim = 0.4cm 0.4cm 0.4cm 0.4cm, scale=0.35, clip=True]{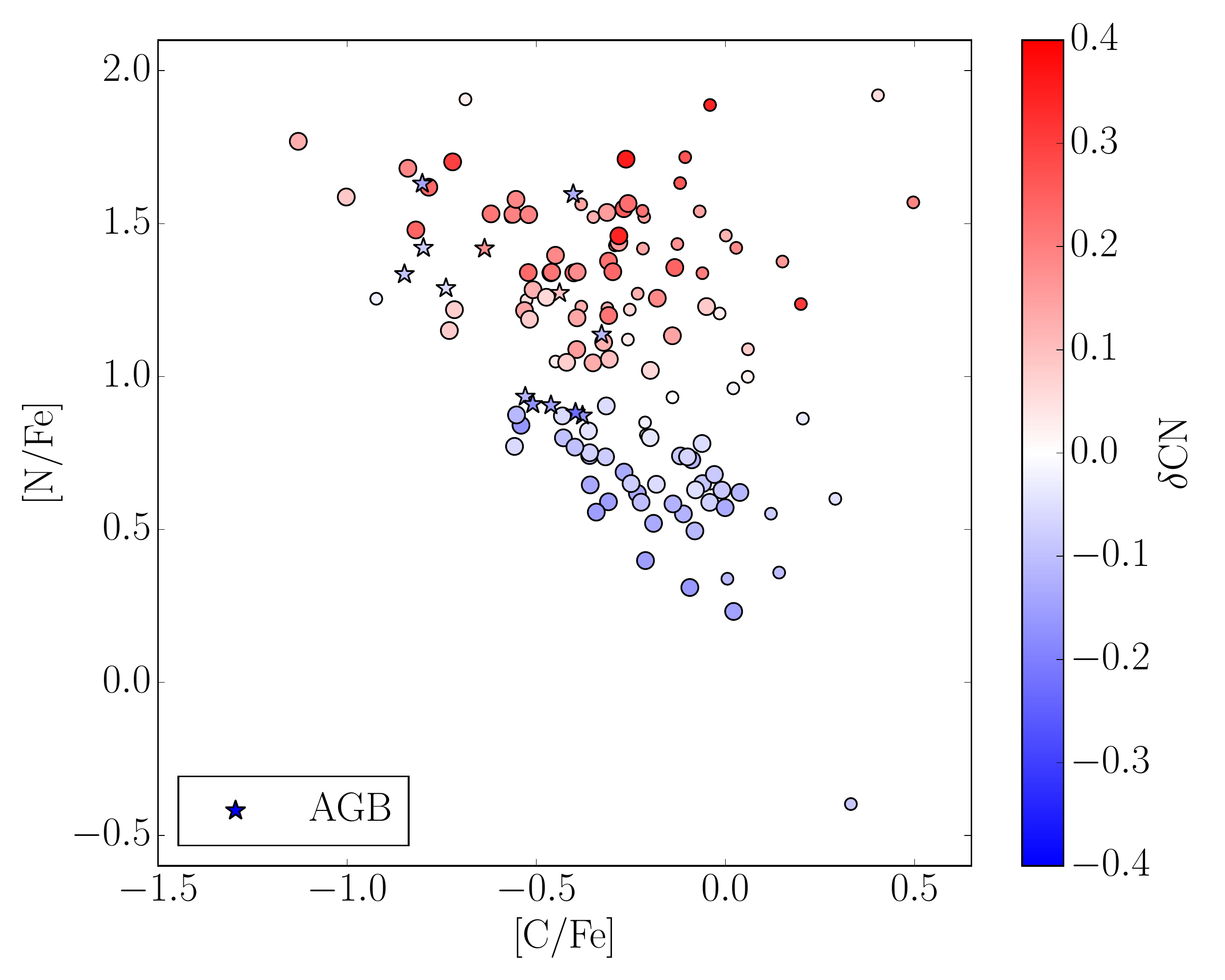}
\caption{[N/Fe] vs. [C/Fe] for all of the members in our sample. Following the same convention as previous figures, AGB stars are indicated as stars. Points are color coded based on their $\delta$CN index as indicated by the color bar on the right side of the plot. Faint stars (stars with $M_v > 1$) are indicated as smaller points compared to brighter stars.}
\label{nfevscfe}
\end{figure}

Our final derived values for [C/Fe] and [N/Fe] are listed in Table \ref{tab:alldata} along with our band measurements and shown in Figure \ref{nfevscfe} where we color code points based on their $\delta$CN values. The anti-correlation between C-N is clearly present, as expected, with N abundances increasing as C abundances decrease. CN-enhanced and CN-normal stars also map directly onto enhanced and normal N abundances, respectively. We also see in Figure \ref{nfevscfe} that RGB stars do not separate cleanly into two populations in the C-N plane, and that the high N stars show a range of C abundances. Two factors contribute to this continuous distribution.

First, the fainter stars in our sample have higher uncertainties than the brighter stars, as explained above. These higher uncertainties ($\sim$0.35 dex for [N/Fe]) cause any boundary between the two populations to become blurred. We have indicated the fainter stars (M$_v > 1$) as smaller points in Figure \ref{nfevscfe}; these points show a more dispersed distribution, particularly populating the region of high N and high C abundances. The faint stars populate this region due to asymmetrical uncertainties in the abundance measurements caused by the band strengths becoming less sensitive to changes in abundance as temperatures increase (see Figure \ref{isolines}). If only the brighter stars are considered (the larger points), the C-N anti-correlation is more pronounced and a somewhat more distinct separation between populations can be seen. 
A second factor, however, contributes to the distribution in Figure \ref{nfevscfe}. The stars sample the entire range of the RGB, and the underlying dependence of abundance on stellar magnitude, with its depletion of C and enhancement of N with evolutionary state, will serve to smooth any initial differences in abundance between the stellar populations. We will discuss these evolutionary effects in a later section (\ref{evolution}).

\section{Results and Discussion} \label{Results}

\subsection{Multiple CN Populations in M10}

\begin{figure}
\centering
\includegraphics[trim = 1.0cm 0.4cm 0.4cm 0.4cm, scale=0.4, clip=True]{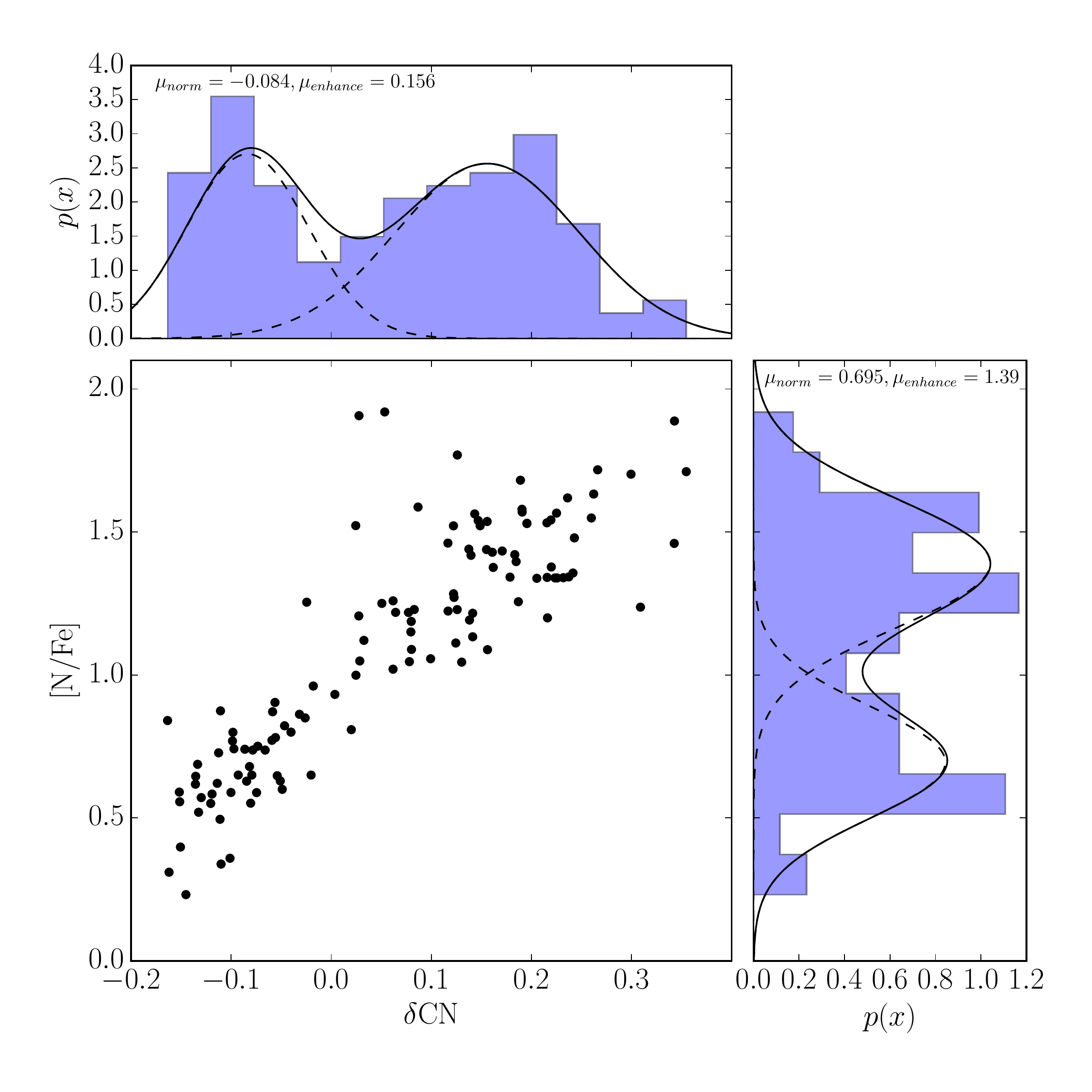}
\caption{[N/Fe] vs. $\delta$CN for the RGB stars in the sample. Probability distributions for the $\delta$CN and [N/Fe] are shown on the x and y axis, respectively. Data are shown in blue with the two Gaussian distributions fit with GMM overplotted. The mean value from each fit is printed at the top of each histogram. $\delta$CN correlates, as expected, with [N/Fe], and both distributions show the clear presence of two distinct populations.}
\label{gmm}
\end{figure}

$\delta$CN measurements have traditionally been used to separate stars into populations in GCs \citep[see][and references therein]{gratton}. Similarly, we used our $\delta$CN index to identify a CN-enhanced and CN-normal population as discussed in Section \ref{bandmeasurements}. Figure \ref{gmm} shows that the $\delta$CN strengths correlate directly with N abundance, as expected. The few outliers at high [N/Fe] for their $\delta$CN strengths are at the extreme magnitude limits of the sample where the CN band is less sensitive to changes in N abundance, and so their $\delta$CN index will be smaller than it is for the bulk of the stars at the same N abundance.

Figure \ref{gmm} also shows the marginal distributions in $\delta$CN and [N/Fe] along the x and y axes, respectively. These two distributions clearly show the presence of several populations. However, it appears that the CN-enhanced population has a broader distribution than the others, suggesting that perhaps there are several populations contributing to this broad peak. Studies have shown that GCs can host more than two populations with an extreme example being the five populations found in NGC 2808 \citep{milone5pop}; \citet{gir,uves} found that in a number of GCs the Na-O anti-correlation shows evidence of an ``extreme" second generation in addition to the ``intermediate" second generation found in all clusters. We used a Gaussian Mixture Model (GMM) to explore how many populations are needed to describe the distributions of both $\delta$CN and [N/Fe].

The GMM fit for the $\delta$CN distribution indicates that the CN-enhanced population does indeed have a larger standard deviation than the CN-normal population, with a standard deviation of 0.092 versus 0.061. We use the Bayesian information criterion (BIC) and the Akaike information criterion (AIC) to compare the success of fitting different models \citep[see e.g.,][]{ivezic}, finding a best fit to both the $\delta$CN and the [N/Fe] distributions with just two populations. Contrary to the case for $\delta$CN, the GMM fit to the [N/Fe] distribution yielded two gaussians with equal standard deviations of 0.2 dex, and a separation of roughly 0.7 dex between their median values. These fits are shown in Figure \ref{gmm}.

The similarity in the range in N abundances in both the CN-normal and CN-enhanced populations indicates that the larger spread in $\delta$CN for the CN-enhanced population is caused by an observational effect of the CN bands rather than an underlying extreme population. \citet{gir,uves} also find that M10 lacks an extreme population in the Na-O anti-correlation. Figure \ref{gmm} suggests that only using CN band strengths to describe characteristics of underlying populations may lead to inferences about the nature of those populations that are not substantiated by elemental abundances.

We find that 75 out of 124 or 60\% of RGB stars in our sample are CN-enhanced based on having a positive $\delta$CN measurement; this agrees with the percentage of second generation stars found based on their Na abundance using the \citet{gir,uves} sample. However, we note that when dividing the stars into bright and faint magnitude bins separated at $M_v = 1$ the result changes slightly. Only 42 out of 80 or 52\% bright RGB stars are CN-enhanced, and 33 out of 44 or 75\% of faint stars are CN-enhanced. Our classification of the fainter stars is much less secure, however, and the percentage of CN-enhanced stars may be more accurately reflected in the brighter sample.

\subsubsection{AGB Stars} \label{AGB}
Recent studies have raised the question of whether there is a lack of CN-enhanced stars on the AGB in GCs relative to the percentage of enhanced stars found on the RGB and MS. \citet{campbell2012} observed a lack of CN-enhanced stars on the AGB of NGC 6752 based on $\delta$CN index measurements, which agreed with what had been determined by \citet{norris} decades prior. Follow up work by \citet{campbell2013} found that the cluster had no AGB stars enhanced in Na abundance compared to the Na distribution of the RGB in NGC 6752, and therefore concluded that the second generation of stars in NGC 6752 had failed to evolve to the AGB from the RGB. A similar study conducted by \citet{maclean2016} found an absence of second generation stars on the AGB for the GC M4 using the Na-O anti-correlation.

In contrast to this work, later studies found a Na-enhanced, second generation of stars along the AGB in other clusters such as 47 Tuc, M13, M5, M3, M2, M4, and NGC 6397 \citep[][]{johnson,garcia,lardo,maclean2018}. \citet{lapenna} conducted new observations of AGB stars in NGC 6752 and found there to be two populations in C-N and Na-O that were not found by \citet{campbell2013} using just the Na distribution. They suggested that the work done by \citet{campbell2013} had a systematic offset in their [Na/Fe] abundances that caused all stars to fall below their threshold set for enhancement. \citet{lapenna} also emphasized that just looking at one abundance distribution can lead to false conclusions about the presence or lack of multiple populations in any sample of stars. Other recent work has proposed that the presence of a second generation of stars on the AGB is dependent on the metallicity and age of the cluster, and finds the presence of a second generation on the AGB to vary from cluster to cluster \citep{wang2016,wang2017}.

From the left panel of Figure \ref{bands}, it is clear that most AGB stars in our sample have normal CN band strengths with only 2 of the 13 ($\sim$15 percent) belonging to the CN-enhanced group. These numbers suggest there is a lack of second generation stars along the AGB as compared to the percentage of stars found to be CN-enhanced in the RGB sample ($\sim$60 percent). However, the CN band is dependent on both the C and N abundance, and the right panel of Figure \ref{bands} also shows that the AGB stars have weak CH band strengths (heavily dependent on C abundance) based on their magnitude. In fact, many of those that appear to be N-normal based on their CN band strength are actually N-enhanced based on their [N/Fe]. This effect can be seen in Figure \ref{nfevscfe} where many AGB stars with high [N/Fe] also have low $\delta$CN values.

When separating populations by N abundance, we find that 8 out of 13 AGB stars are N-enhanced, matching the percentage seen in the RGB. This serves as another reminder that while the CN band strength is useful for separating stars into multiple populations, many effects can change the CN band strength such as surface temperature, pressure, and C abundance. These other factors need to be considered fully before using CN as a way to determine the population of a star. The fundamental basis for identifying multiple populations rests on the abundances of the stars, which is why using the N abundances to determine populations is more accurate. We also note that our classification of RGB stars based on CN band strength rather than N abundance is still secure due to the tight correlation between $\delta$CN and [N/Fe] shown in Figure \ref{gmm}. 

\subsection{Spatial Distribution of Multiple Populations}

Our large sample size and broad field coverage from 0.06 half-light radii out to $\sim$6 half-light radii (with a majority of stars within $\sim$4 half-light radii) allows us to study the spatial distributions of the two populations we have identified in this cluster. All formation models \citep[see e.g.,][]{dercole2008,decressin2007a,decressin2007b,bekki2010} agree that second generation stars should form more centrally concentrated than first generation stars. Subsequent dynamical evolution will gradually erase these initial differences in the spatial distribution \citep[see e.g.,][]{vesperini2013} eventually leading to the complete spatial mixing of the two populations. Depending on the cluster's evolutionary phase some memory of the initial spatial distribution of the two populations might be preserved and indeed in several clusters \citep[see e.g.,][]{bellini2009,lardo2011,hbtemp,m13ona,milone2012,cordero2014,simioni2016} second generation stars have been found to be more centrally concentrated than first generation stars.

\begin{figure}
\centering
\includegraphics[trim = 0.2cm 0.4cm 0.4cm 0.2cm, scale=0.46, clip=True]{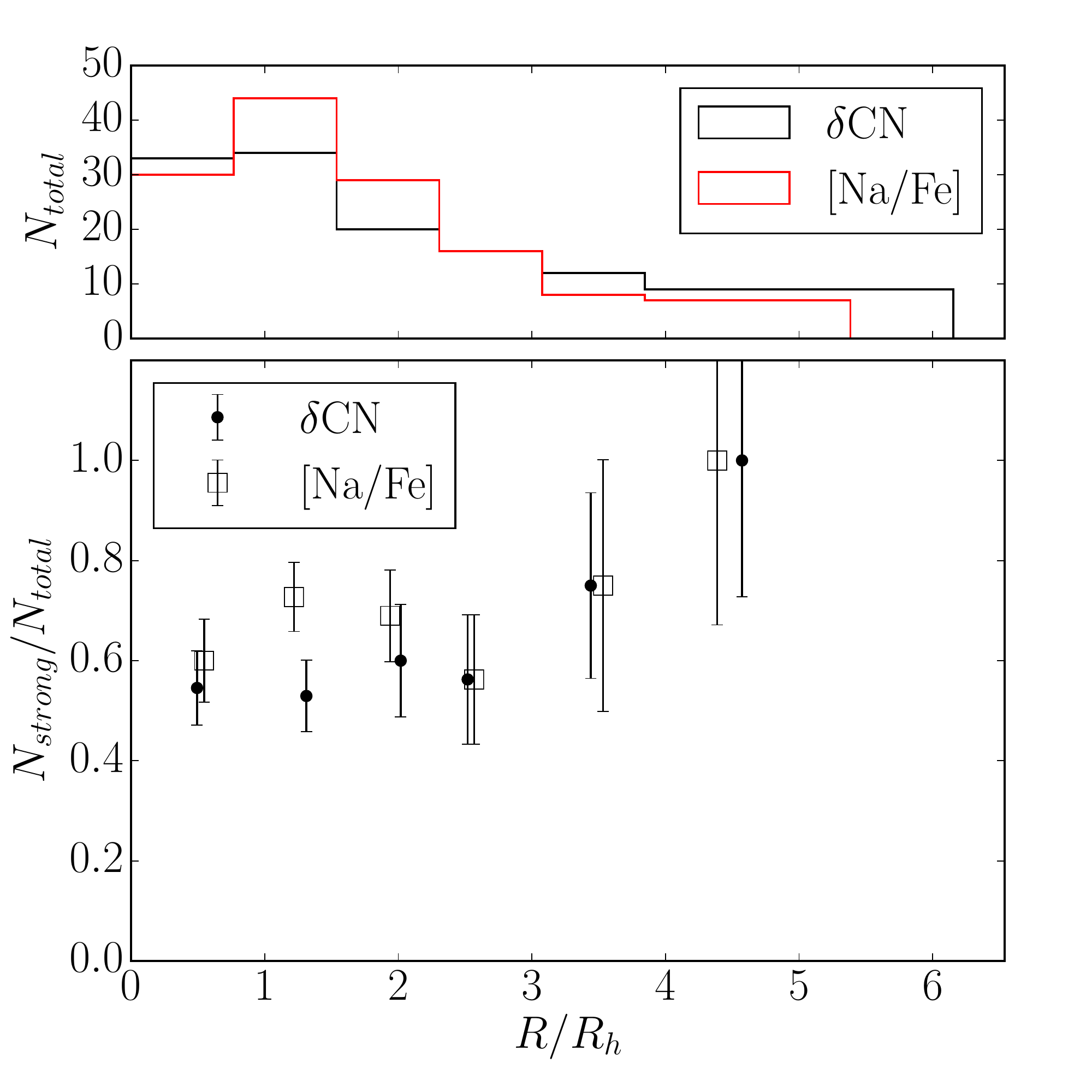}
\caption{The radial distribution of the ratio of the number of CN or Na-enhanced, second generation stars to the total number of stars. The top panel shows a histogram of the total number count in each sample with our $\delta$CN data shown in black and the Na data shown in red. Bin sizes are 1.5 arcmin out to 4 half-light radii. All stars at a greater radius were combined into one bin for each sample. Na abundances are taken from \citet{gir,uves}. The points shown in the bottom panel are centered at the average position of the stars in a given radial bin as shown in the histogram above. The populations overall are roughly equal with a slight enhancement in the second generation (60 percent of stars in the total sample). Both populations appear to have similar radial distributions.}
\label{radial}
\end{figure}

\citet{vesperini2013} have found that in order to reach complete spatial mixing a cluster must be in its advanced evolutionary stage and have lost at least 60-70 \% of its initial mass due to the effects of two-body relaxation \citep[see also][]{miholics2015}. An indication that M10 might indeed have undergone a significant mass loss due to two-body relaxation and is dynamically old comes from the study of the cluster's stellar mass function and large degree of mass segregation \citep{beccari2010,webb2017}. These dynamical indicators suggest that the two populations should have completely mixed and are consistent with our findings reported in Figure \ref{radial}. This figure shows the radial profile of the ratio of second generation stars to the total number of stars. Irrespective of whether the identification of second generation is based on our determination of CN or the Na abundance from \citet{gir,uves} we find no evidence of a significant radial gradient confirming the expectation that M10 is a dynamically evolved cluster in which the two populations have reached complete spatial mixing \citep[see also][for other examples of mixed clusters]{dalessandro2014,nardiello2015,cordero2015}.

\subsection{Comparison to Other Methods of Identifying Multiple Populations}

\subsubsection{Comparison with Na-O Anti-Correlation} 

In addition to the C-N anti-correlation, the Na-O anti-correlation has been often used in the literature to identify multiple populations in GCs \citep[see e.g.,][and references therein]{gir,uves,gratton}. \citet{smith2013} and \citet{smith2015a,smith2015b} have compared the classification of multiple populations based on Na and O abundances with that based on N abundance and tested whether N-enhanced stars do indeed coincide with those characterized by O depletion and Na enhancement. Their analysis showed that in the three clusters they studied (47 Tuc, M71, and M5) stars classified as second generation by an Na overabundance and an O depletion are also characterized by a CN enhancement. They find that while the correlation holds true overall, one star in M5 and a few stars in 47 Tuc appear to be CN-enhanced, but do not have the Na enhancement to place them in the second generation based on their Na abundance. Similar evidence of this effect has also been observed in M53 by \citet{bobergm53}.

\begin{figure*}
\plottwo{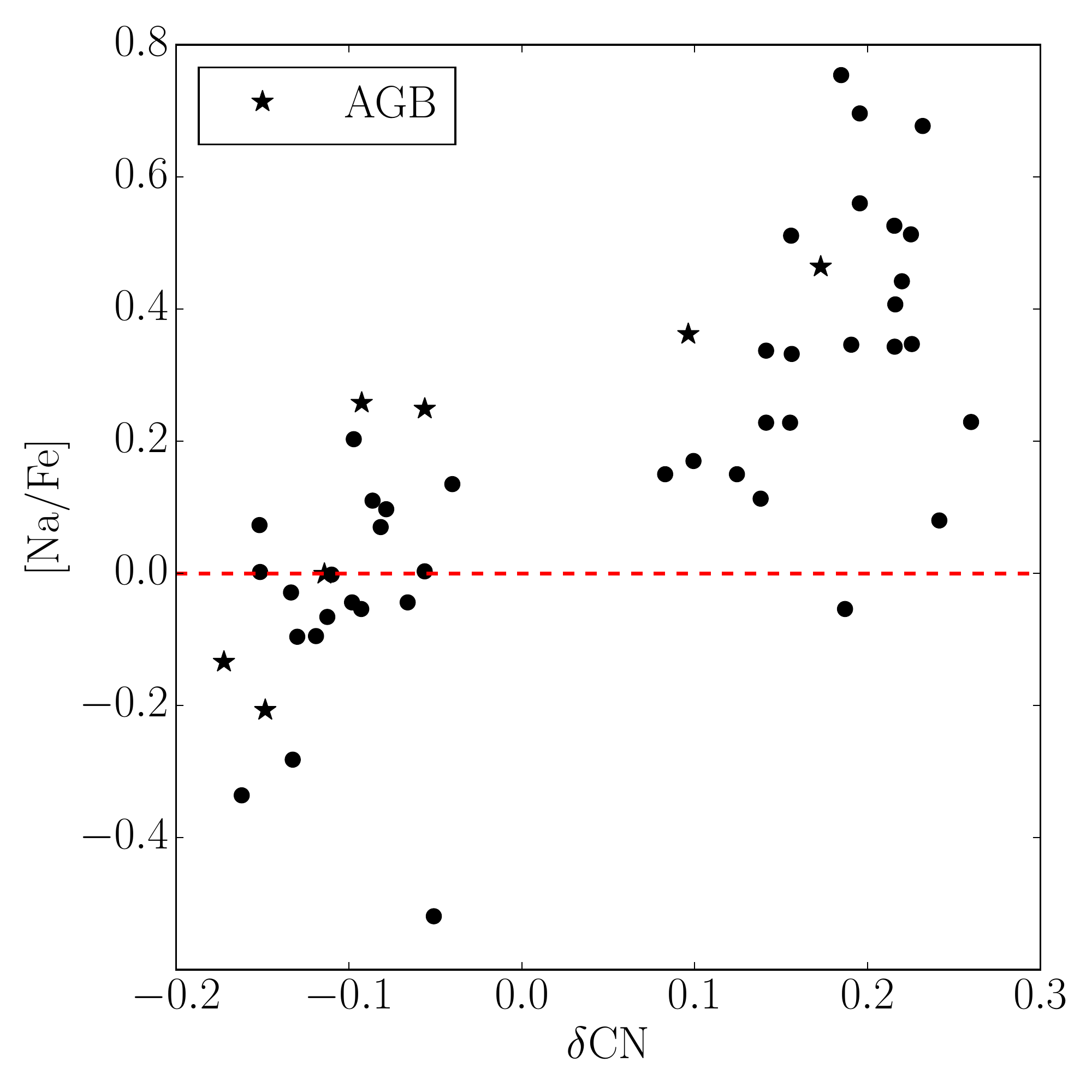}{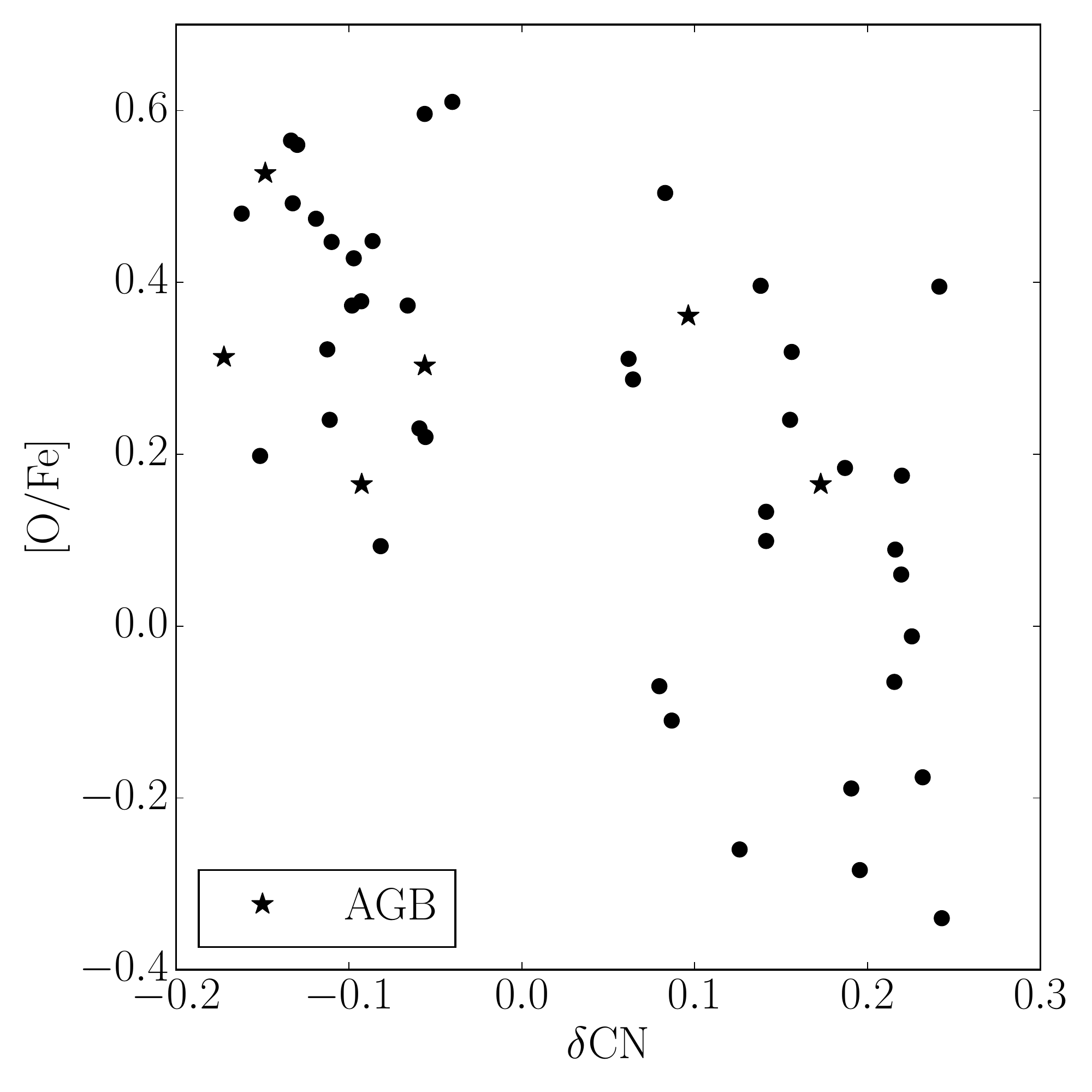}
\plottwo{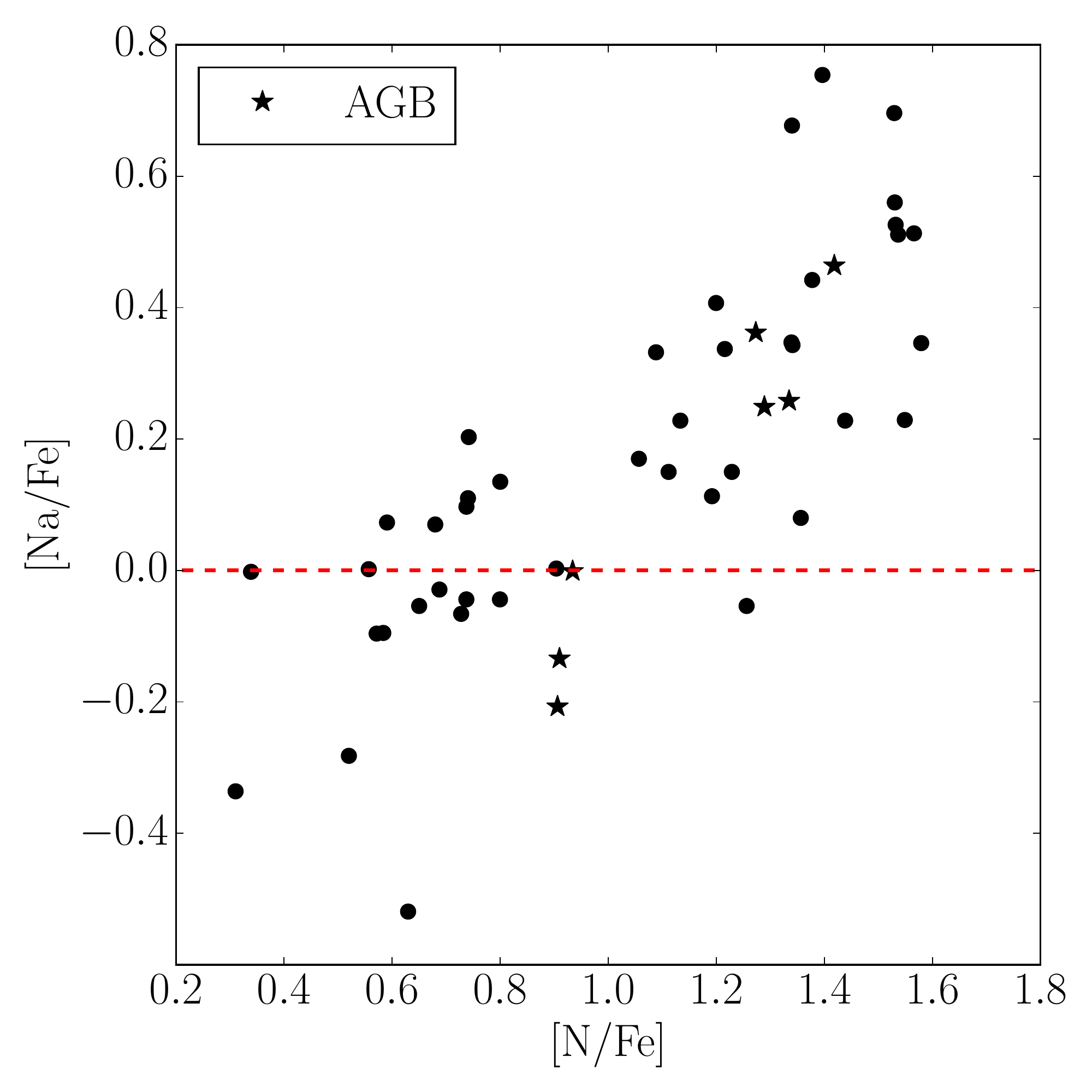}{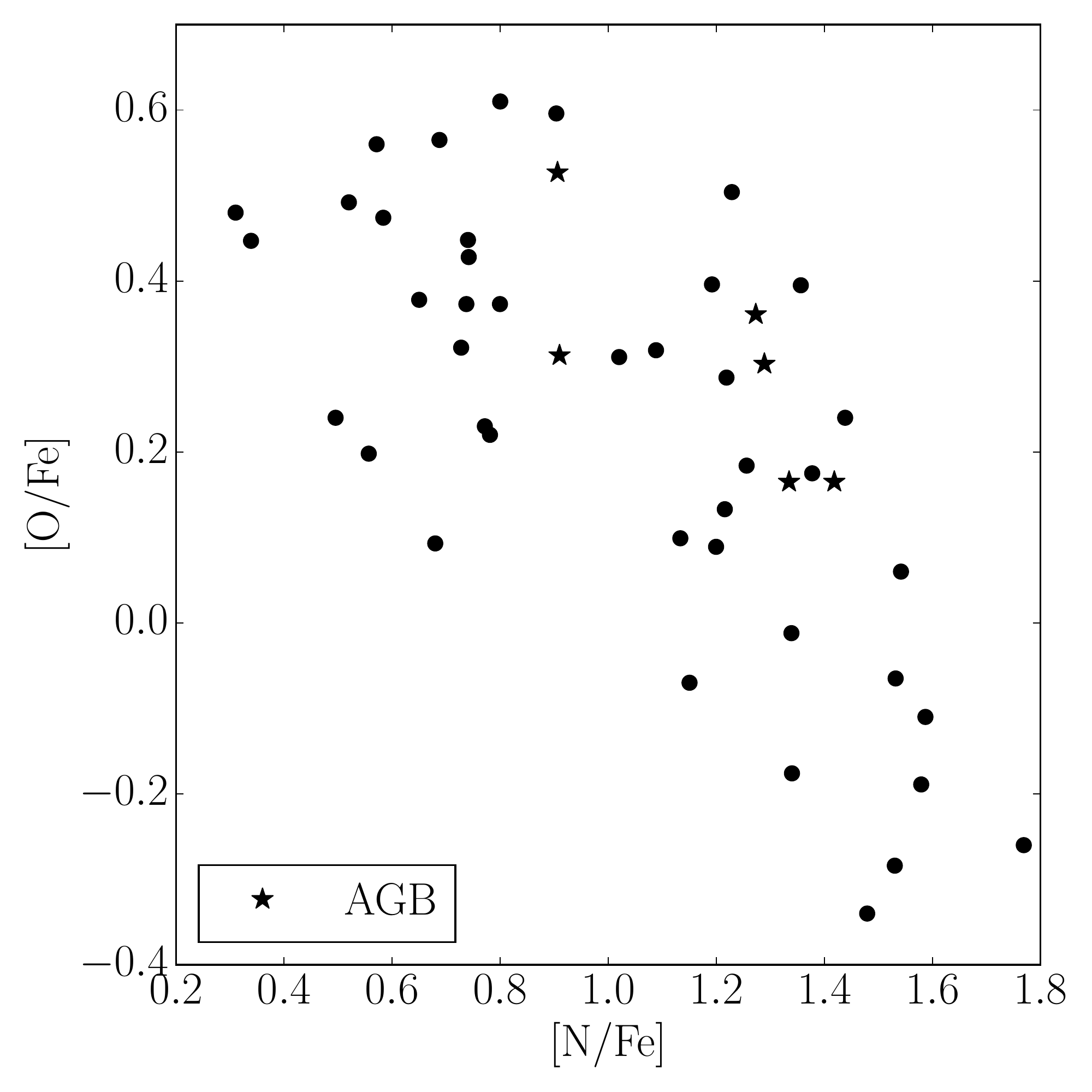}
\caption{\textbf{Top:} [Na/Fe] and [O/Fe] vs $\delta$CN (left and right, respectively). Na and O measurements are from \citet{gir,uves}. The red dashed line in the top left plot indicates the separation between populations in [Na/Fe] as defined by \citet{gir,uves}. $\delta$CN = zero defines the CN-normal and enhanced populations. AGB stars are indicated as stars and RGB stars as circles. \textbf{Bottom:} Same as above vs. [N/Fe]. The two AGB stars with enhanced Na but weak $\delta$CN have shifted into the correct group as these stars actually have enhanced [N/Fe].}
\label{nafe}
\end{figure*}

We studied the same relation between CN band strength and Na abundance to see if these stars that have depleted Na but enhanced CN are seen in M10. The top left plot of Figure \ref{nafe} shows the [Na/Fe] measured by \citet{gir,uves} for stars in our sample vs. $\delta$CN. Also shown is the line that \citet{gir,uves} use to separate the Na-enhanced and normal populations in M10, which is defined as 4 standard deviations above the minimum Na abundance observed in the cluster. There is a general correlation of $\delta$CN strength with [Na/Fe] as expected if both methods are identifying the same populations (i.e., CN-enhanced are Na-enhanced and vice versa). Similarly, the top right plot of Figure \ref{nafe} shows the $\delta$CN measurements also anti-correlate with [O/Fe] as expected. There is also a noticeable clear separation between the $\delta$CN values of the two populations while the Na abundance shows a more continuous distribution. While we find one star to be slightly under abundant in Na compared to its $\delta$CN value, uncertainties in the [Na/Fe] measurements of $\sim$0.1 dex \citep{gir,uves} could easily place it in the Na-enhanced group. We do not see any stars with the under-abundance in Na seen in 47 Tuc, which is 0.2-0.4 dex below the average Na for a CN-enhanced star. 

We also find a group of stars that are slightly enhanced in Na even though they are considered CN-normal, which was also observed in 47 Tuc, M71, and M5 by \citet{smith2013} and \citet{smith2015a,smith2015b}. We note, however, that the two of these stars that are most enhanced in Na compared to other CN-normal stars are AGB stars, which reflect the phenomenon discussed in Section \ref{AGB}. In the lower left panel of Figure \ref{nafe}, we plot the [Na/Fe] vs [N/Fe] where we see that these AGB stars are in fact N-enhanced, and [N/Fe] correlates very well with [Na/Fe]. The bottom right panel of Figure \ref{nafe} also shows that [O/Fe] anti-correlates with [N/Fe] as expected with no evidence of the anomalous stars found in the clusters studied by \citet{smith2013}, \citet{smith2015a,smith2015b}, and \citet{bobergm53}.

\subsubsection{Comparison with HST UV Legacy Archive Photometry}
Recent work uses HST UV photometry to find populations using a pseudocolor that is dependent on the C, N, and O abundances of a star \citep[see][]{piotto,miloneatlas}. The pseudo-color enhances the separation between the RGBs of the populations in the cluster. Stars in different populations with different light element abundances such as C, N, and O, alter the OH, NH, CN, and CH bands that appear in the regions of the HST filters used. As expected, stars rich in N appear on the ``bluer" branch and stars depleted in N appear on the ``redder" branch.

Most notably, this technique has been used to identify up to five different populations in the GC NGC 2808 \citep{milone5pop}. M10 was included in the HST UV Legacy Survey \citep{miloneatlas} which showed the RGB of M10 split into two sequences, consistent with two populations in the cluster. \citet{miloneatlas} also determined that M10 was roughly 64 percent second generation. These findings match the results we find from our CN band study as well as the results from the Na-O study by \citet{gir,uves}.

\begin{figure}
\centering
\includegraphics[trim = 0.4cm 0.4cm 0.4cm 0.4cm, scale=0.46, clip=True]{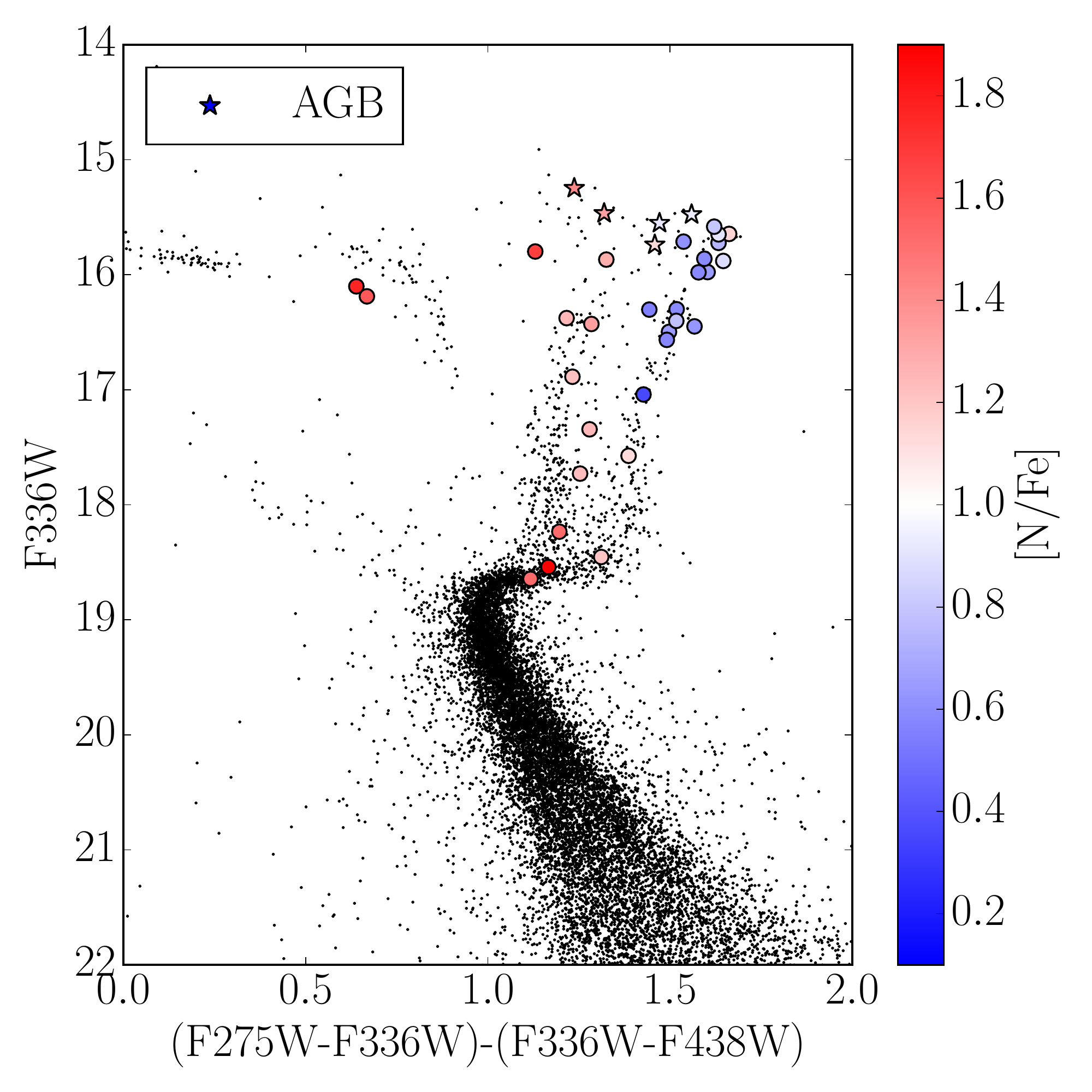}
\caption{A pseudo-color magnitude diagram \citep{piotto} to determine multiple populations photometrically shown with stars color-coded based on [N/Fe]. RGB stars are shown as circles and AGB stars are stars. The pseudo-color enhances the separation between the red giant branches of the populations in the cluster. As expected, N-enhanced stars appear on the ``bluer" branch and stars weak in nitrogen appear on the ``redder" branch.}
\label{hst}
\end{figure}

We make a direct comparison between our CN method of identifying multiple populations with the pseudocolor method described above. While our data cover a much wider field than the HST photometry, our sample includes 35 stars observed by HST and are shown in Figure \ref{hst}, the pseudo-color magnitude diagram for M10. From the figure, we see that all stars identified as second generation by being N-enhanced also fall along the bluer RGB sequence in the pseudoCMD for M10. We also find that almost all N-normal stars fall along the correct redder RGB sequence for the first generation. There are two very ``blue" stars in Figure \ref{hst} that do not fall along either RGB sequence. These stars are actually the two reddest stars in our sample based on B-V color, and therefore have very low UV flux, which causes their ``blue" position in the pseudocolor magnitude diagram. Based on these results, we find that the two methods of identification are consistent and classify stars into the same populations.

\subsection{Effect of Evolution on C and N Abundances} \label{evolution}

C and N abundances change in the evolved stars of GCs as they ascend the RGB. Specifically, low mass stars go through a second period of deep mixing after the first dredge up that decreases the C abundance while simultaneously increasing the N abundance.
This deep mixing refers to a non-canonical evolutionary event in low mass stars that occurs on the red giant branch just after the star has gone past the LF bump. In these stars, CN(O)-cycle material is dredged up to the surface of the star from the hydrogen burning envelope, which causes changes in surface abundance. 
The two leading theories to explain this mixing are a thermohaline process where the chemical weight becomes inverted due to the final step in the pp chain that converts two $^3$He atoms to a $^4$He and two protons \citep{eggleton2006,eggleton2008,charbonnel}, and an effect due to stellar rotation \citep{sweigart,chaname,palacios}. Either way, a non-canonical mechanism needs to be put into place to explain how material is able to cross over the radiative zone from the hydrogen burning envelope into the convective zone of the star.

\begin{figure*}
\centering
\includegraphics[trim = 0.4cm 0.4cm 0.4cm 0.4cm, scale=0.45, clip=True]{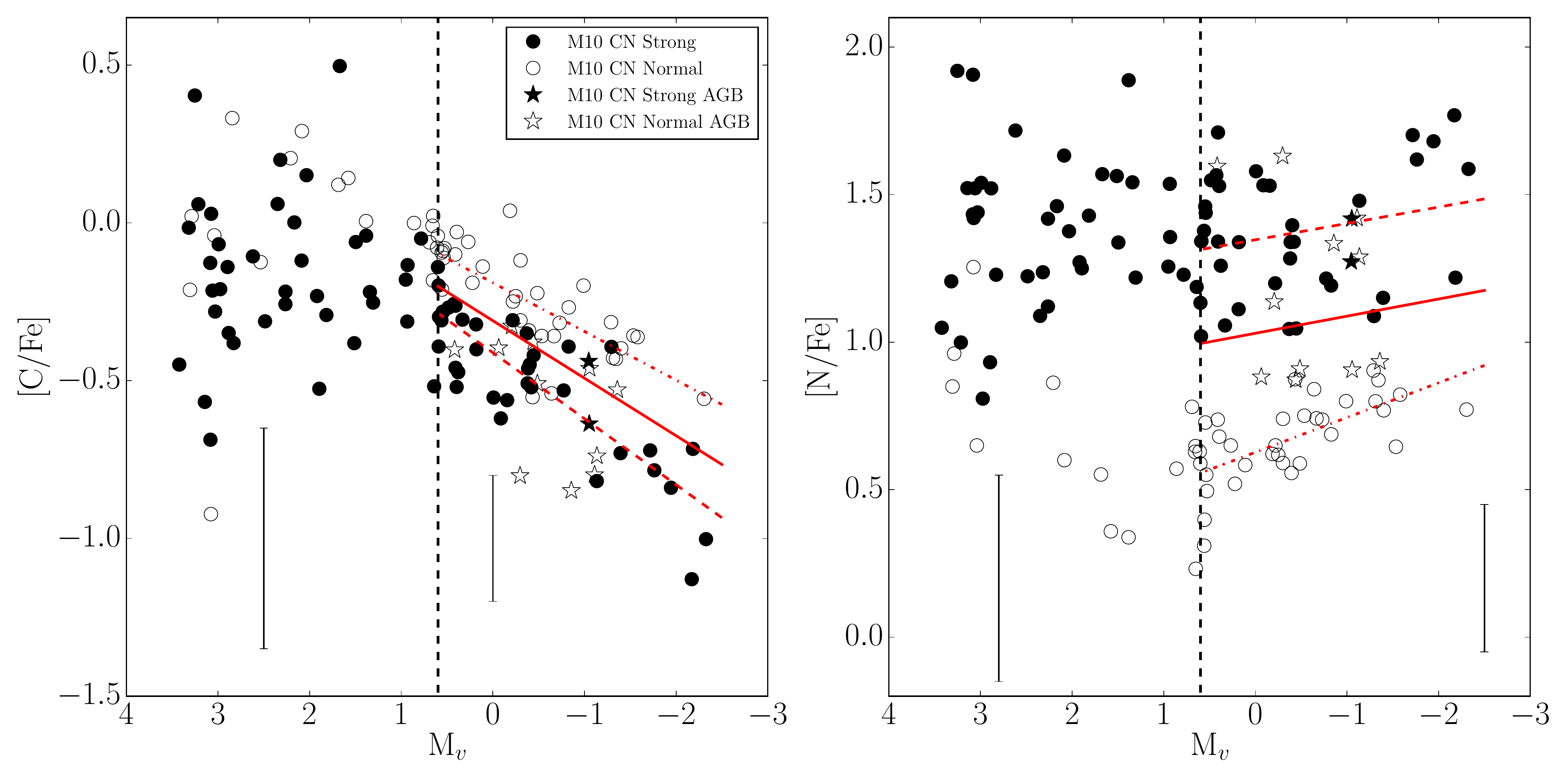}
\caption{[C/Fe] and [N/Fe] vs. M$_v$ for all 137 member stars in M10. Stars are coded based on their $\delta$CN with CN-enhanced stars as filled circles and CN-normal stars as open circles. AGB stars are shown as stars with CN-enhanced filled and CN-normal open. A dashed line at $M_v = 0.7$ represents the LFB \citep{lfb}, where secondary mixing is expected to begin for this cluster. Linear fits to RGB stars are shown in red for CN-normal, enhanced, and both populations as dot dashed, dashed, and solid lines, respectively. Representative error bars are shown for stars that are brighter and fainter than $M_v = 1$.}
\label{xfevsm}
\end{figure*}

While some efforts have been made to constrain the rate of this mixing observationally (e.g., \citealt{martellmixing}), unknown parameters such as the initial carbon abundance of the stars have prevented these methods from determining the mixing rate in clusters with high precision. However, our large sample size allows us to determine mixing rates through direct fits to the C abundance as a function of magnitude from a homogenous sample rather than the heterogeneous samples of \citet{m3cfe}. In fact, there are enough stars in the sample that the two populations can be fit independently without having to make assumptions about the initial C abundance. We use a simple linear-least squares regression that begins at the LF bump measured to be M$_{bump}$ $\sim$~0.7 by \citet{lfb}. Fits were made to both [C/Fe] and [N/Fe] as a function of magnitude as both N and C will be affected by the deep mixing that brings CN(O)-cycle material to the surface of the star. We show these fits (discussed below in more detail) in Figure \ref{xfevsm}.

\begin{figure}
\centering
\includegraphics[trim = 0.4cm 0.4cm 0.4cm 0.4cm, scale=0.35, clip=True]{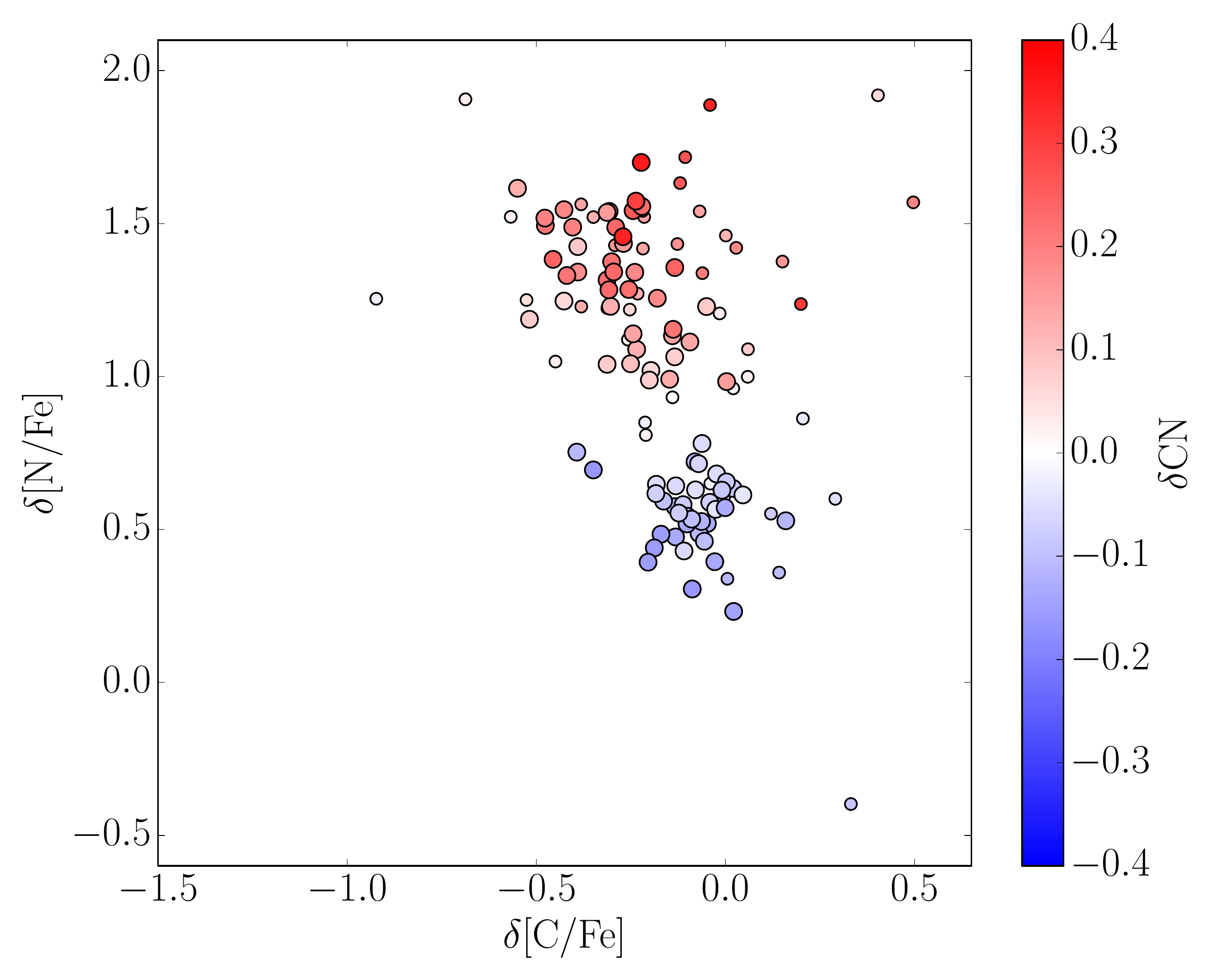}
\caption{$\delta$[N/Fe] vs. $\delta$[C/Fe] created by removing the trends found with magnitude, which are shown in Figure \ref{xfevsm}. The color convention is the same as that used in Figure \ref{nfevscfe}.}
\label{deltanfevsdeltacfe}
\end{figure}

In Section \ref{candn}, we discussed the magnitude dependence on C and N that is present in the C-N anti-correlation shown in Figure \ref{nfevscfe}. The deep mixing phenomenon that depletes C while enhancing N as stars evolve up the giant branch alters the initial abundances of the respective populations in the cluster. We can use the linear fits to [C/Fe] and [N/Fe] versus magnitude to correct for these evolutionary effects and create normalized $\delta$[C/Fe] and $\delta$[N/Fe] measures for each star. These are shown in Figure \ref{deltanfevsdeltacfe}. Comparing Figures \ref{nfevscfe} and \ref{deltanfevsdeltacfe} shows that the anti-correlation has become even clearer, and the two populations separate quite cleanly in the C-N plane. The fainter stars in the sample show a slightly more scattered distribution, consistent with their larger uncertainties.

In Figure \ref{xfevsm}, we show the [C/Fe] and [N/Fe] vs magnitude for all of the members we measured in M10; CN-normal stars are shown as open points and CN-enhanced stars as filled. The LFB for the cluster is shown as a black dashed line. The resulting fits for C and N are presented as a red solid line for all RGB stars, a dotted-dashed line for the CN-normal population, and a dashed line for the CN-enhanced population. While the fainter stars show a larger spread due to their uncertainties, the overall abundance stays constant fainter than the LFB; we find no evidence of pre-RGB bump mixing as reported by \citet{angelou2012}. For magnitudes brighter than the LFB, the carbon abundance begins to decrease in both populations, while the nitrogen abundance begins to increase. 

For [C/Fe], we find a slope of 0.21 $\pm$ 0.02 dex mag$^{-1}$ for the CN-enhanced population, and a slope of 0.15 $\pm$ 0.03 dex mag$^{-1}$ for the CN-normal. For [N/Fe], we find slopes of -0.06 $\pm$ 0.06 dex mag$^{-1}$ and -0.12 $\pm$ 0.06 for the CN-enhanced and normal populations, respectively. These slopes are listed in Table \ref{tab:rates}. We also find that the initial values for [C/Fe] and [N/Fe] predicted by these fits match well within uncertainties with the measurements for the fainter stars in our sample. While there are no other C or N measurements of stars this faint for M10 in the literature, our [C/Fe] values do agree with stars in a similar magnitude range in M13, a cluster of similar metallicity \citep{briley2002}.

We note that the slopes for the change in [C/Fe] and the change in [N/Fe] for each of the two populations are consistent within the uncertainties. The change in [C/Fe] as a function of magnitude matches the inverse change of [N/Fe] for the CN-normal stars, which is expected since the CN(O)-cycle is the cause of the changes in abundance. While the change in [N/Fe] does not match the change in [C/Fe] for the CN-enhanced population, this difference is also expected. As explained in \citet{angelou2012}, the change in the overall N abundance is continuous, but the rise in the measured [N/Fe] value is slow when the value of [N/Fe] is already high due to it being a logarithmic ratio. However, the difference (while small) in slopes of [C/Fe] between the two populations could indicate that stars in the second generation are mixing at a somewhat increased rate compared to the first generation. We discuss possible reasons for this in the section below.

 \begin{deluxetable}{ccc}
\tabletypesize{\footnotesize}
\tablecolumns{3}
\tablewidth{0pt}
\tablecaption{\\Rate of Abundance Change vs. Magnitude
\label{tab:rates}}
\tablehead{
\colhead{\textbf{Sample}} & \colhead{\textbf{$d$[C/Fe]/$dM_v$}} & \colhead{\textbf{$d$[N/Fe]/$dM_v$}}}
\startdata
\textbf{M10} & &\\
\hline
All & 0.18 $\pm$ 0.03 & -0.06 $\pm$ 0.06\\ 
CN-enhanced & 0.21 $\pm$ 0.02 & -0.06 $\pm$ 0.06\\
CN-normal & 0.15 $\pm$ 0.03 & -0.12 $\pm$ 0.06\\
\hline
\textbf{M13$^{1}$} & &\\
\hline
All & 0.43 $\pm$ 0.05 & \nodata \\
CN-enhanced & 0.40 $\pm$ 0.05 & \nodata \\
\hline
\textbf{M3$^{2}$} & & \\
\hline
All & 0.22 $\pm$ 0.03 & \nodata \\
\enddata
\tablecomments{1. \citet{m13cfe}, 2. \citet{m3cfe}}
\end{deluxetable}

\subsubsection{Comparison to Other Clusters}

To further understand the evolutionary effects on the C and N abundances in M10, we compare our data to similar data sets for M3 measured by \citet{m3cfe} and for M13 as compiled by \citet{m13cfe} from various literature sources. These clusters were chosen because they are relatively well studied, which gives comparably sized data sets, and because of their similar age and metallicity to M10. M3 and M13 have often been compared to one another as they embody the classic second parameter problem in GCs \citep[e.g.,][]{caloi}.

\begin{figure}
\centering
\includegraphics[trim = 0.4cm 0.4cm 0.4cm 0.4cm, scale=0.43, clip=True]{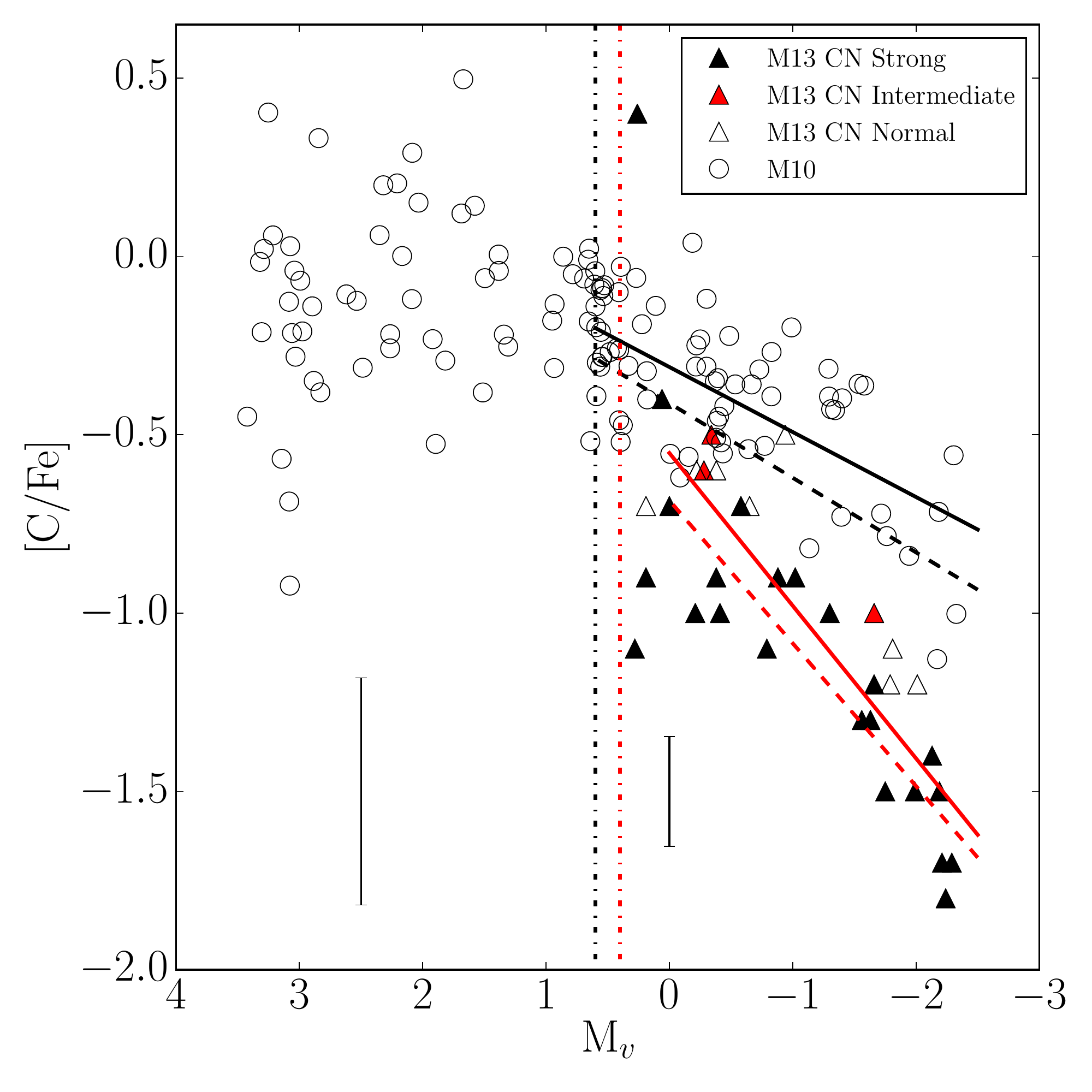}
\caption{[C/Fe] vs. M$_v$ for RGB stars from Figure \ref{xfevsm} shown with data from M13 \citep{m13cfe} shown as triangles. M13 are coded as CN-enhanced (filled), $\delta$CN index (open), and CN intermediate (red) as determined by \citet{m13cfe}. All M10 points are shown as open circles for clarity. The black and red dot dashed lines indicate the LFB for M10 and M13, respectively \citep{lfb}. The black lines show fits to the M10 RGB stars and red lines are fits M13 RGB stars. Fits to the entire sample are shown as solid lines, and fits to CN-enhanced stars are shown as dashed lines for each cluster.}
\label{cfecompare}
\end{figure}

M3 is found to have a very similar carbon depletion rate to M10 with \citet{m3cfe} finding a value of 0.22 $\pm$ 0.03 dex mag$^{-1}$ overall, while we find a value of 0.18 $\pm$ 0.03 dex mag$^{-1}$ for M10. However, M13 is found to have a greater depletion rate of roughly double that of M10. Figure \ref{cfecompare} shows the data for M13 from \citet{m13cfe} plotted over our data for M10 for comparison. The LFB for each cluster is shown as a dot dashed line \citep{lfb}. 

We evaluated linear fits to [C/Fe] as a function of absolute magnitude for the M13 data as we did for M10. These fits are shown in Figure \ref{cfecompare} as a red line for M13 and a black line for M10. Solid lines indicate a fit to the entire sample of stars for a cluster, while dashed lines indicate fits to the CN-enhanced sub-sample. A fit was not made to the CN-normal population of M13 because only 3 stars were available from the \citet{m13cfe} sample. While our expectation was for the mixing rate to be similar for M13 as that of M3 and M10 due to their similar metallicities, it appears from our fits that M13 reduces its carbon much faster than M10; the total M13 sample has a slope of 0.43 $\pm$ 0.05 dex mag$^{-1}$, while the M10 sample has a slope less than half of that at 0.18 $\pm$ 0.03 dex mag$^{-1}$. This difference suggests that the astrophysical process behind the mixing is dependent on a factor/s other than metallicity that may be different between the two clusters. One of the possibilities for this factor is a higher He abundance in M13 than M10.

%Several indicators suggest this additional factor is a difference in He abundance between the two clusters.
One piece of evidence for a higher He abundance in M13 is the difference in the Na-O anti-correlation based on data from \citet{gir,uves}, \citet{m13ona}, and \citet{m13cfe}. While both M10 and M13 show an anti-correlation in Na-O as expected, the anti-correlation is more extended in M13 than in M10. \citet{hbtemp} quantify the extension of the Na-O anti-correlation by looking at the interquartile range (IQR) of the [O/Na] values for each cluster; more extended anti-correlations have a higher IQR([O/Na]). Based on Na and O measurements compiled from various sources by \citet{m13cfe} and measured by \citet{m13ona} for M13, and \citet{gir,uves} for M10, we find that M13 has an IQR([O/Na]) of 0.70 compared to 0.56 for M10. \citet{hbtemp} also find that the IQR([O/Na]) correlates with the maximum temperature reached on the horizontal branch. This correlation combined with the higher IQR([O/Na]) for M13 indicates that stars reach higher surface temperatures on the horizontal branch of M13 than the same stars in M10. Because stars with higher surface temperatures on the horizontal branch tend to have higher helium abundances, stars in M13 likely have a higher helium abundance.

Other indications that M13 may have a higher He abundance than other clusters of similar metallicity come from photometric studies. M13 has a slightly bluer horizontal branch than M3 \citep{caloi,denissenkov2017}, consistent with higher temperatures and higher He abundance. A recent study of the RGB bumps of multiple populations also suggests a larger He spread in M13 than M3 \citep{lagioia2018}. \citet{lfb} have shown that the LFB of M13 is slightly brighter than that of M10; the LFB has been shown in studies such as \citet{milone5pop} to correlate with He abundance.

While other factors such as the C+N+O abundance affect these photometric indicators and may complicate cluster comparisons, the majority of evidence is consistent with M13 having a higher He abundance than M10. If so, then the enhanced mixing rate seen in M13 may be influenced by its higher He abundance as this would be one of the few differences between the two clusters.

\section{Conclusions} \label{Conclusions}
We have measured CN and CH band strengths for over 124 RGB stars and 13 AGB stars in the GC M10 using low-resolution spectroscopy. We used these measurements to identify CN-enhanced and CN-normal populations. C and N abundances were then calculated for all stars in the sample. Below is a list of our conclusions based on this analysis:
\begin{enumerate}
\item M10 clearly shows two populations in both the CN band and [N/Fe] distributions for stars from the tip to the base of the red giant branch. Gaussian mixture modeling shows that while the CN-enhanced population has a larger range in $\delta$CN compared to the CN-normal, the two populations have similar dispersions in [N/Fe] comparable to observational errors. These two populations are centered at [N/Fe]=0.70 and 1.4 dex with standard deviations of roughly 0.2 dex. 
\item When compared to other methods of identifying multiple populations such as the Na-O anti-correlation \citep{gir,uves} and HST UV photometry \citep{miloneatlas}, we find that the CN band analysis classifies stars in common to these samples into the same populations. We do not find any of the outlying stars with enhanced CN and normal Na that have been discovered in M71, 47 Tuc, and M53 by \citet{smith2015a}, \citet{smith2015b}, and \citet{bobergm53}, respectively.
\item Our study of the spatial distributions of the two populations identified in M10 does not reveal any dependence of the fraction of CN-enhanced stars on the distance of the cluster center. A similar conclusion is reached if second generation stars are identified by using the Na-O data from \citet{gir,uves}. The mixing of the two populations is consistent with the expectations based on this cluster's dynamical evolution.
\item The RGB stars in M10 clearly show an anti-correlation between [C/Fe] and [N/Fe] as seen in other GCs and expected for the cluster. The distribution in the C-N plane is continuous, not clearly identifying each population. However, by accounting for the evolutionary changes to C and N as the stars ascend the RGB, the two populations separate clearly.
\item While there appear to be fewer AGB stars that are CN-enhanced than expected based on the percentage seen in the RGB sample, we find that this is a consequence of the low C abundance of these stars. Since these stars have already been through deep mixing, they have lower C abundances (and therefore, lower CH and CN band values) on average than RGB stars. When classifying the AGB stars by their [N/Fe] values, however, the N-enhanced stars make up 60\%, consistent with the RGB ratio.
\item The evolution of [C/Fe] and [N/Fe] as a function of M$_v$ has been studied for each population to determine a rate of change above the LF bump. We find that in M10 the carbon depletion rates between the second and first populations agree within uncertainties. While M3 has a similar rate of depletion, we find that M13 depletes carbon much faster than M10. Because these clusters are all of similar metallicity, the differing depletion rates are likely caused by another factor, which may be the He abundance.
\end{enumerate}

\section{Acknowledgements}
We would like to thank Roger A. Bell for making the SSG program available to us. We would also like to thank Zachary Maas for his help in obtaining the stellar spectra during our 2016 run at WIYN. Additionally, we would like to thank Dianne Harmer, Daryl Willmarth, and all of the observing assistants who helped with our multiple runs on Hydra. 

This publication makes use of data products from the Two Micron All Sky Survey, which is a joint project of the University of Massachusetts and the Infrared Processing and Analysis Center/California Institute of Technology, funded by the National Aeronautics and Space Administration and the National Science Foundation.

\end{document}